\documentclass{imsart}
\RequirePackage{natbib}
\usepackage{amsmath,amssymb,amsthm}
\usepackage{graphics,epsfig}
\usepackage{hyperref}
\usepackage{natbib}
\usepackage{color}
\usepackage{graphicx}
\usepackage{caption}
\usepackage{subcaption}
\usepackage{float}
\usepackage{dsfont}
\usepackage{mathrsfs}
\usepackage{multirow}
\usepackage{comment}
\usepackage{bm}

\def \qq {\boldsymbol{q}}

\def \ra {\rightarrow}

\def \E {\mathbb{E}}

\def \a {\alpha}

\newtheorem{example}{\bf Example}

\newtheorem{definition}{\bf Definition}
\newtheorem{defn}[definition]{\bf Definition}

	\newtheorem{theorem}{\bf Theorem}
	\newtheorem{thm}[theorem]{\bf Theorem}
	\newtheorem{prop}{\bf Proposition}
	\newtheorem{lem}[theorem]{\bf Lemma}
	\newtheorem{as}{\bf Assumption}

\renewcommand{\epsilon}{\varepsilon}

\begin{document}

\begin{frontmatter}

\title{Optimization of a Dynamic Profit Function using Euclidean Path Integral}
\runtitle{Euclidean Path Integral}

\begin{aug}

\author{\fnms{Paramahansa} 
	\snm{Pramanik}
	\ead[label=e1]{ppramanik1@niu.edu}}
\and
\author{\fnms{Alan M.} 
	\snm{Polansky}
	\ead[label=e2]{polansky@niu.edu}}
	
\runauthor{P. Pramanik and A. M. Polansky}

\affiliation{Northern Illinois University}

\address{Department of Statistics and Actuarial Science \\
	DeKalb, IL 60115 USA}
\end{aug}
  
\begin{abstract}
A Euclidean path integral is used to find an optimal strategy for a firm under a Walrasian system, Pareto optimality and a non-cooperative feedback Nash Equilibrium. We define dynamic optimal strategies and develop a Feynman type path  integration method to capture all non-additive convex strategies. We also show that the method can solve the non-linear case, for example Merton-Garman-Hamiltonian system, which the traditional Pontryagin maximum principle cannot solve in closed form. Furthermore, under Walrasian system we are able to solve for the optimal strategy under a linear constraint with a linear objective function with respect to strategy.
\end{abstract}



\end{frontmatter}

\section{Introduction}
In this paper we consider dynamic profit maximization over a time interval with finite horizon $t>0$. The objective is to find an optimal strategy for a firm in a system whose state dynamics are specified by a  stochastic differential equation. The instantaneous profit function we consider depends on the time $s$, a real-valued measure of the market share of the firm $x(s)$, and the real-valued dynamic strategy of the firm $u(s)$. The profit function is represented by  $\pi[s,x(s),u(s)]\in\mathbb{R}$. Here $x\in \mathcal{X}$ and $ u\in \mathcal{U}$, where $\mathcal{X}$ is a functional space corresponding to the set of all market share trajectories and $\mathcal{U}$ is a functional space corresponding to the set of all possible strategies available to the firm. We assume the functional spaces $\mathcal{X}$ and $\mathcal{U}$ are bounded and complete. The profit over the time interval  $[0,t]$ is measured by the stochastic integral 
\[
\int_0^t\pi[s,x(s),u(s)]ds,
\]
using the It\^o representation of the integral \citep{oksendal2003}.
The dynamics of the market share are  given by 
\begin{equation}\label{mkt}
	dx(s)=\mu[s,x(s),u(s)]ds+\sigma[s,x(s),u(s)]dB(s),
\end{equation}
where $B(s)$ is Brownian motion process. 

In this paper we are interested in calculating three types of equilibria:  
Walrasian, Pareto and Nash. 
The Walrasian system is a fundamental market structure in economics, and is the basis of many other market systems \citep{walras1900}. The main assumption under this system is that each firm is small when compared to the entire industry and therefore does not influence the industry price. The industry consists of all the firms, and its price is determined by the entire system. In this system, a single firm can earn at least zero profit in the long run. Therefore, if a single firm  wants to survive it has to achieve its average cost \citep{walras1900}.

\begin{definition}\label{DWalrus}
	The continuous path of market share  $x^*(s)\in X$ and a continuous set of optimal strategies  $u^*(s)\in U$ constitute a Walrasian Equilibrium  if for every time point $s\in[0,t]$,
	\begin{equation}\label{w}
	\E \int_{0}^t \pi[s,x^*(s),u^*(s)] ds\geq\E \int_0^t  \pi[s,x(s),u(s)] ds,
	\end{equation}
	with market dynamics defined in Equation (\ref{mkt}).
\end{definition}

Definition \ref{DWalrus} implies that each firm under the Walrasian system faces identical market dynamics. In this case, finding the optimal strategy of a firm corresponds to solving the optimization problem
\begin{align}\label{n1}
&\max_{u \in U}\ \Pi(u,t)=\max_{u \in U}\ \E\ \int_{0}^{t}\ \pi[s,x(s),u(s)]ds,
\end{align} 
under the constraint given in Equation (\ref{mkt}), and initial condition $x(0)=x_0$.

Determining Pareto and Nash equilibria requires us to consider the other firms in the industry. Suppose that there are $k$ firms in an economy, where the strategy function of firm $\rho$ is given by $u_\rho(s)$ for $\rho=1,..., k$, $u_\rho\in \mathcal{U}_\rho\subset\mathcal{U}$, where $\mathcal{U}_\rho$ is the set of all available strategies of firm $\rho$, and $\mathcal{U}$ is the set of all available strategies in the market. Let $x_\rho(s)$ be the measure of market share for firm $\rho$. Let $\mathbf{x}(s)$ and $\mathbf{u}(s)$ be the vectors containing the elements $x_\rho(s)$ and $u_\rho(s)$ for $\rho=1,...,k$, respectively. Each firm has a dynamic profit function  $\pi_\rho[s,\mathbf{x}(s),\mathbf{u}(s)],$ 
with market dynamics specified by
\begin{equation}\label{vectordynamics}
d\mathbf{x}(s)=\bm{\mu}[s,\mathbf{x}(s),\mathbf{u}(s)]ds+\bm{\sigma}[s,\mathbf{x}(s),\mathbf{u}(s)]d\mathbf{B}(s),
\end{equation}
where $\bm{\mu}[s,\mathbf{x}(s),\mathbf{u}(s)]$ is an $k$-dimensional drift function, $\bm{\sigma}[s,\mathbf{x}(s),\mathbf{u}(s)]$ is an $k\times m$-dimensional diffusion function, and $\mathbf{B}(s)$ is an $m$-dimensional Brownian motion process. 
The initial condition is $\mathbf{x}(0)=\mathbf{x}_0\in\mathbb{R}^k$.

Pareto optimality is an economic environment where each player benefits at the expanse of the other players \citep{greenwald1986,mas1995}. Therefore, Pareto optimality insures the greatest mutual benefit for all of the players simultaneously. Mathematically this is equivalent to maximizing the total dynamic profit, 
\[
\overline{\Pi}_\text{P}(u,t)=
E\int_0^t\sum_{\rho=1}^k\a_\rho\pi_\rho[s,\mathbf{x}(s),\mathbf{u}(s)]ds,
\]
where $\a_\rho$ is the profit weight corresponding to $\rho^{th}$ firm such that $\sum_{\rho=1}^k\a_\rho=1$.

\begin{definition}\label{paretodef}
	The strategies $\mathbf{u}^*\in \mathcal{U}^k$, constitute a cooperative Pareto Equilibrium for the $\rho^{th}$ firm if 
	\begin{equation}\label{p}
	\E \int_{0}^t\ \sum_{\rho=1}^k\ \a_\rho\pi_\rho[s,\mathbf{x}(s),\mathbf{u}^*(s)]\ ds\geq \E \int_0^t\ \sum_{\rho=1}^k\ \a_\rho\pi_\rho[s,\mathbf{x}(s),\mathbf{u}(s)]\ ds,
	\end{equation}
	for $\rho=1,...,k$ subject to the Equation (\ref{vectordynamics}) with initial condition $\mathbf{x}(0)=\mathbf{x}_0$, where $\a_\rho$ is the profit weight of $\rho^{th}$ firm such that
	\[
	\sum_{\rho=1}^k \a_\rho=1.
	\]
\end{definition}

Assuming $\pi_\rho[s,\mathbf{x}(s),\mathbf{u}(s)]$ is non-negative and differentiable, Fubini's Theorem implies that the cooperative Pareto equilibrium the optimization problem for the $\rho^{th}$ firm is
\begin{equation}\label{n6}
\max_{u_\rho\in \mathcal{U}}\overline{\Pi}(\mathbf{u},t)=
\max_{u_\rho\in U} \int_0^t\left\{\E\sum_{\rho=1}^k\a_\rho
\pi_\rho[s,\mathbf{x}(s),\mathbf{u}(s)]\right\}ds,
\end{equation}
subject to Equation (\ref{vectordynamics}), with initial condition $\mathbf{x}(0)=\mathbf{x}_0$. In other words, Equation (\ref{n6}) implies that $\rho^{th}$ firm performs its optimization in light of the optimal strategies of the other firms. 

\begin{definition}
	In the non-Cooperative feedback Nash framework
	a set of optimal strategies $\mathbf{u}^*(s)$ form a non-cooperative feedback Nash equilibrium if 
	\begin{align}
	\E\left\{\int_0^t\pi_\rho[s,\mathbf{x}^*(s),\mathbf{u}^*(s)]ds\right\}\geq
	\E\left\{\int_0^t\pi_\rho[s,\mathbf{x}(s),\hat{\mathbf{u}}_\rho(s)]ds\right\},\notag
	\end{align}
	for all $\rho\in\{1,...,k\}$ where $t\in(0,\infty)$,
	subject to the constraints,
	\begin{align}\label{n11}
	dx^*(s)&=\bm{\mu}[s,\mathbf{x}^*(s),\mathbf{u}^*(s)]ds+\bm{\sigma}[s,\mathbf{x}^*(s),\mathbf{u}^*(s)]dB(s),
	\end{align}
	\begin{align}\label{n11.1}
	dx_\rho(s)&=\mu[s,x_\rho(s),\tilde{\mathbf{u}}^*_\rho(s)]ds+\sigma[s,x_\rho(s),\tilde{\mathbf{u}}^*_\rho(s)]dB(s), 
	\end{align}
	and $\mathbf{x}(0)=\mathbf{x}_0$, for $\rho=1,..., k$, where 
	\[\tilde{\mathbf{u}}^*_\rho(s)=[u_1(s),...,u_{\rho-1}(s),u_\rho^*(s),u_{\rho+1}(s),\ldots,u_k(s)]',\] 
	and 
	\[\hat{\mathbf{u}}^*_\rho(s)=[u_1^*(s),...,u^*_{\rho-1}(s),u_\rho(s),u_{\rho+1}^*(s),\ldots,u_k^*(s)]'.\]
	\label{def3}
\end{definition}

Hence, firm $\rho$ has the optimization problem 
\begin{equation}\label{n9}
\max_{u_\rho\in U}\tilde{\Pi}(u_\rho,t)=
\max_{u_\rho\in U}\E\int_0^t\pi_\rho[s,\mathbf{x}(s),\hat{\mathbf{u}}_\rho(s)]ds,
\end{equation}
subject to the constraints in Equations (\ref{n11}) and (\ref{n11.1}) and initial conditions $\mathbf{x}(0)=\mathbf{x}^*(0)=\mathbf{x}_0$.

Traditionally, these optimization problems are solved by using the Pontryagin principle \cite{pontryagin1987} after solving the Hamilton-Jacobi-\\Bellman equation. See \cite{bellman1952,bellman2013,bellman2015,ljungqvist2012,pontryagin1966,stokey1989} and \cite{yeung2006}. The main problem with this method is that finding a solution often requires obtaining a complicated value function.
An alternative method for solving optimal control problems is based on principles from quantum mechanics and path integrals. These methods have previously been used in motor control theory  \citep{kappen2005,theodorou2010,theodorou2011}, and finance \citep{belal2007}. There are three mathematical representations of this approach based on partial differential equations, path integrals, and stochastic differential equations  \citep{theodorou2011}. Partial differential equations give a macroscopic view of an underlying physical process, while path integrals and stochastic differential equations give a more microscopic view. Furthermore, the Feynman-Kac formula yields a special set of Hamiltonian-Jacobi-Bellman equations which are backward parabolic partial differential equations \citep{kac1949}. Only a few problems in finance are directly tractable by Pontryagin maximum principle and solving the Hamiltonian-Jacobi-Bellman equation usually involves solving a  system of differential equations which is often a difficult task.  The potential advantage of the quantum approach is that a general non-linear system, such as Merton-Garman Hamiltonian, can be impossible to solve analytically. The quantum method allows a different approach to attack these problems and sometimes can give simplified solutions \citep{belal2007}. Path integrals are widely used in physics  as a method of studying stochastic systems. In finance, path integrals have been used to study the theory of options and interest rates \citep{linetsky1997,lyasoff2004}. A rigorous discussion of the application of different types of quantum path integrals in finance is given in \cite{belal2007}. The idea is that, in quantum mechanics a particle's evolution is random. This is analogous to the evolution of a stock price having non-zero volatility. 

Motivated by \cite{belal2007} we consider a firm's real-valued measure of market share 
as a stochastic process and use the principles of path integral as the basis for our mathematical model. The assumption is that since a firm is a very small part of an industry and  an economy, and is subject to many small stochastic perturbations, the movement of its share  will behave like a quantum particle in physics. 
Although these methods have been used in quantum approaches to financial problems we are not 
aware of their use in stochastic optimization problems for the economic systems studied here.

\section{Main results}

Define a non-negative measurable discounted profit function for a single firm as 
\[
\pi[s,x(s),u(s)]=\exp(-\zeta s)\tilde{\pi}[s,x(s),u(s)].
\]
Assume that $\pi$ is a finite $C^\infty$ function with respect to $ x(s)$ and $u(s)$ where $\zeta\in[0,1]$ is a constant discount rate of profit over $s\in[0,t]$. The function $\tilde{\pi}[s,x(s),u(s)]$ is the actual profit at time $s$, and is assumed to be quadratic in terms of change in time, non-decreasing in output price, non-increasing in input price, homogeneous of degree one in output and input prices, convex in output and input prices, continuous in output and input prices, and is continuous with respect to $s$. We assume that $x(s)$ is a time dependent measure of a stochastic market dynamic and the strategy $u(s)$ is a deterministic function of $x$. Further technical assumptions are given in the Appendix.

To optimize the dynamic profit function $\Pi$ defined in Definition \ref{DWalrus}  
with respect to the strategy $u$ we need to specify a function $g:[0,t]\times\mathcal{X}\rightarrow\mathbb{R}$ to favor strategies that respect the dynamics specified by Equation (\ref{mkt}). In the standard Lagrangian framework this function is specified as $g(s,x)=\lambda[dx(s)-h(s,x)]$, where $h$ is a function that specifies the dynamics of the system and $\lambda$ is the Lagrange multiplier. 

\begin{prop}[Walrasian Equilibrium]\label{p1} 
	An optimal strategy for maximizing the dynamic profit function $\Pi(u,t)$ with respect to the control $u$ and constraint  
	\[
	dx(s)=\mu[s,x(s),u(s)]\ ds+\sigma[s,x(s),u(s)]\ dB(s),
	\]
	with initial condition $x(0)=x_0$ is the solution of the equation 
	\begin{multline}\label{o}
	\left[\frac{\partial}{\partial u}f(s,x,u)\right]
	\left[\frac{\partial^2}{\partial x^2}f(s,x,u)\right]^2\\=
	2\left[\frac{\partial}{\partial x}f(s,x,u)\right]
	\left[\frac{\partial^2}{\partial x\partial u}f(s,x,u)\right],
	\end{multline}
	with respect to $u$ as a function of $x$ and $s$ evaluated at $x=x(s)$, where  
	\begin{multline}\label{w11}
	f(s,x,u)=
	\pi(s,x,u)+g(s,x)+\frac{\partial}{\partial s}g(s,x)\\+
	\mu(s,x,u)\ \frac{\partial}{\partial x}g(s,x)+
	\mbox{$\frac{1}{2}$}\sigma^2(s,x,u) \
	\frac{\partial^2}{\partial x^2} g(s,x).
	\end{multline}	
\end{prop}

\begin{example}\label{ex2}
	Suppose that a firm under a Walrasian system has the objective function 
	\[ 
	\E\left\{\int_0^t\exp(-\zeta s)[px(s)-cx(s)u(s)]ds\right\},
	\]
	where $\zeta\in(0,1]$ is a constant discount rate over time interval $[0,t]$, $p>0$ is constant price, $x(s)$ is the total output, a twice differentiable function of $s$, $c$ is a positive constant marginal cost, and $u$ is the total expenditure on advertising. 
	Consider market dynamics given by 
	\begin{equation}\label{dynamic1}
	dx(s)=[ax(s)-u(s)]ds+\sqrt{\sigma x(s)u(s)}dB(s),
	\end{equation}
	where $a$ and $\sigma$ are two positive and finite constants. The negative terms in the drift part of Equation (\ref{dynamic1}) and the objective function reflect the firm's cost of advertising its product as its strategy. The diffusion component of Equation (\ref{dynamic1}) reflects the amount of variation in the system. To apply Proposition \ref{p1} we specify $g(s,x)$ to represent the market dynamics. For a fixed positive Lagrangian multiplier $\lambda^*$ let $g(s,x)=\lambda^* s[ax^2-b]$ where $b$ is a positive number such that $a<b$. Equation (\ref{w11}) yields 
	\[
	f(s,x,u)=x\exp(-\zeta s)(p-cu)+s\lambda^*(ax^2-b)+2\lambda^*sax(ax-u)+\lambda^*\sigma asux.
	\]
	Therefore 
	\[
	\frac{\partial}{\partial x}f(s,x,u)= 
	\exp(-\zeta s)(p-cu)+2\lambda^*sax+2\lambda^*as(ax-u)+2\lambda^*sa^2x +\lambda^*\sigma asu,
	\]
	\[
	\frac{\partial}{\partial u}f(s,x,u)=
	-cx\exp(-\zeta s)-2\lambda^*asx+\sigma asx\lambda^*=A(s,x),
	\]
	\[
	\frac{\partial^2}{\partial x^2}f(s,x,u)=2\lambda^*as(1+2a)=B(s),
	\]
	and
	\[
	\frac{\partial^2}{\partial x\partial u}f(s,x,u)=
	-c\exp(-\zeta s)-2\lambda^*as+\sigma\lambda^*as=D(s). 
	\]
	Equation (\ref{o}) then implies that an optimal Walrasian strategy for this system is given by
	\begin{equation}\label{e3}
	\phi_w^*(s,x)=\frac{1}{\sigma\lambda^*as-2\lambda^*as-c\exp(-\zeta s)}
	\left[\frac{A(s,x)B^2(s)}{2D(s)}-E(s,x)\right],
	\end{equation}
	where $E(s,x)=p\exp(-\rho s)+2\lambda^*asx+4\lambda^*a^sx$, $D(s)\neq0$, 
	$\sigma\lambda^*as\neq2\lambda^*as+c\exp(-\rho s)$ and 
	\[
	\frac{A(s,x)B^2(s)}{2D(s)}-E(s,x)\neq 0.
	\]
\end{example}

In Example \ref{ex2} both the objective function and the market dynamics are linear continuous mappings from strategy space to the real line. According to the Generalized Weierstrass Theorem there exists an optimal strategy. One such strategy is given in Equation (\ref{e3}). The Pontryagin maximum principle cannot be used to find a closed-form optimal strategy for this system.

\begin{example}\label{ex3}
	Suppose that a firm under the Walrasian system produces consumer goods with objective 
	function
	\[
	\E\left\{\int_0^t\exp(-\zeta s)\left[R(x)-cu^2\right]ds\right\},
	\]
	where $\zeta\in(0,1]$ is a constant discount rate over time interval $[0,t]$, $R(x)$ is the total revenue function such that it can be multiplicatively separable by $d^2/ds^2$ as discussed in the Appendix, $c$ is the constant cost multiplied by squared strategy function $u(s)$. The main difference between this example with Example \ref{ex2} is that, the strategy $u(s)$ is a $C^2$ function and hence, we can calculate optimal strategy using Pontryagin's maximum principle. 
	Assume the market dynamics of the firm follow
	\begin{equation}\label{dynamic2}
	dx(s)=[bx(s)-u(s)]ds+\sqrt{2 b x(s)}dB(s),
	\end{equation}
	where $b$ is a positive constant. 
	We will use our method and the traditional Pontryagin maximum principle to find the optimal strategy of this Walrasian firm under a consumer good industry.
	
	As the consumption of consumer goods increases exponentially, a Walrasian firm under this sector should face the market dynamics which shows the behavior in Equation (\ref{dynamic2})  \citep{cohen2004, remus2019}. 
	Assume for a fixed Lagrangian multiplier $\lambda^*$ the $g(s,x)$ function is an exponential function with the trend of Equation (\ref{dynamic2}).
	That is $g(s,x)=\lambda^*\exp(sbx-d)$. 
	Equation (\ref{w11}) yields
	\[
	f(s,x,u)=\exp(-\zeta s)\left[R(x)-cu^2\right]+g(s,x)[1+bx+sb^2x(1-b)-sbu].
	\]
	Therefore
	\begin{eqnarray*}
		\frac{\partial}{\partial x}f(s,x,u)
		& = & \exp(-\zeta s)
		\frac{\partial}{\partial x}R(x)+
		g(s,x)[b+sb^2(1-b)]\\ & &+
		\frac{\partial}{\partial x}g(s,x)
		\{1+bx[1+sb(1-b)]\}-
		sbu\frac{\partial}{\partial x}g(s,x) \\
		& = & A_0(s,x)-sbu
		\frac{\partial}{\partial x}g(s,x),
	\end{eqnarray*}
	\[
	\frac{\partial}{\partial u}f(s,x,u)=
	-\left[2cu\exp(-\zeta s)+sbg(s,x)\right],
	\]
	\begin{eqnarray*}
		\frac{\partial^2}{\partial x^2}f(s,x,u) & = & 
		\exp(- \zeta s)
		\frac{\partial^2}{\partial x^2}R(x)+ 
		\frac{\partial^2}{\partial x^2}g(s,x)
		\{1+bx[1+sb(1-b)]\} \\
		& & +2\frac{\partial}{\partial x}g(s,x)
		[b+sb^2(1-b)]-sbu\mbox{$\frac{\partial^2}{\partial x^2}$}g(s,x) \\
		& = & A_1(s,x)-sbu
		\frac{\partial^2}{\partial x^2}g(s,x), 
	\end{eqnarray*}
	and 
	\[
	\frac{\partial^2}{\partial x\partial u}f(s,x,u)
	=-sb\frac{\partial}{\partial x}g(s,x),
	\]
	where 
	\begin{multline*}
	A_0(s,x)=\exp(-\zeta s)\mbox{$\frac{\partial}{\partial x}$}R(x)+g(s,x)\left[b+sb^2(1-b)\right]\\+\mbox{$\frac{\partial}{\partial x}$}g(s,x)\left\{1+bx[1+sb(1-b)]\right\}
	\end{multline*}
	 and
	\begin{multline*}
	A_1(s,x)=\exp(-\zeta s)\mbox{$\frac{\partial^2}{\partial x^2}$}R(x)+\mbox{$\frac{\partial^2}{\partial x^2}$}g(s,x)\left\{1+bx[1+sb(1-b)]\right\}\\+2\mbox{$\frac{\partial}{\partial x}$}g(s,x)\left[b+sb^2(1-b)\right]
	\end{multline*} 
	 Equation (\ref{o}) yields a cubic strategy function $u$ such that,
	\[B_0(s,x) u^3+B_1(s,x) u^2+B_2(s,x) u+B_3(s,x)=0,\]
	with
	\[
	B_0(s,x)=2c(sb)^6\exp(-\zeta s)g^2(s,x),
	\]
	\[
	B_1(s,x)=
	s^2b^3g(s,x)[s^5b^4g^2(s,x)-4c\exp(-\zeta s)A_1(s,x)],
	\]
	\[
	B_2(s,x)=
	2\left\{c\exp(-\zeta s)A_1^2(s,x)-s^3b^4g^2(s,x)\left[A_1(s,x)-2\right]\right\},
	\]
	\[
	B_3(s,x)=
	sbg(s,x)\left[A_1^2(s,x)-2sbA_0(s,x)\right],
	\]
	and the optimal Walrasian strategy becomes,
	\begin{multline*}
	\phi_w^*(s,x)=
	D_1(s,x)+\left\{D_2(s,x)+\left[D_2^2(s,x)+\left(D_3(s,x)-D_1^2(s,x)\right)^3\right]^{\frac{1}{2}}\right\}^{\frac{1}{3}}\\+\left\{D_2(s,x)-\left[D_2^2(s,x)+\left(D_3(s,x)-D_1^2(s,x)\right)^3\right]^{\frac{1}{2}}\right\}^{\frac{1}{3}},
	\end{multline*}
	such that \[D_1(s,x)=-\frac{B_1(s,x)}{3B_0(s,x)},\] \[D_2(s,x)=D_1^3(s,x)+\frac{B_1(s,x)B_2(s,x)-3B_0(s,x)B_3(s,x)}{6B_0^2(s,x)},\] \[D_3(s,x)=\frac{B_2(s,x)}{3B_0(s,x)}\] and $B_0(s,x)\neq 0$. The important part of this result is 
	that we start with a $g(s,x)$ function such that it is a $C^2$ 
	function within $[0,t]$ and we get the optimal strategy by 
	solving a cubic equation. 
	
	For comparison, Walrasian optimal strategy under Pontryagin 
	maximum principle is found by \cite{yeung2006} as 
	$\phi_w^*(s,x)=0$ or 
	\[
	\phi_w^*(s,x)=
	\frac{bx\exp(-\zeta s)}
	{\exp(-\zeta s)
		\left[1+\mbox{$\frac{1}{2}$}\exp(\zeta s)\right]}.
	\]
\end{example}

\begin{example}\label{ex4}
	Suppose that a pure Walrasian firm in the consumer goods 
	industry has the objective function 
	\[
	\E\left\{\int_0^t\exp(-\zeta s)[R(x)-cu^2]ds\right\},
	\]
	where $\zeta\in[0,1]$ is a constant discount rate over $[0,t]$, $R(x)$ is the total revenue function, and $c$ is the constant cost multiplied by squared strategy function $u(s)$. As we assume the the firm is pure Walrasian, the market dynamics it faces does not depend on the strategy and has the form
	\[dx(s)=bx(s)\ ds+\sqrt{\sigma x(s)}\ dB(s),\]
	where $b$ and $\sigma$ are two positive constants. 
	
	For the Quantum approach assume $g(s,x)=\lambda^*\exp(sbx)$
	for a fixed Lagrange multiplier $\lambda^*$
	\citep{cohen2004,remus2019}. 
	Equation (\ref{w11}) yields
	\[
	f(s,x,u)=\exp(-\rho s)\big[R(x)-cu^2\big]+g(s,x)\left[1+sb^2x\left(1+\mbox{$\frac{1}{2}$}s\sigma\right)\right].
	\]
	Therefore,
	\begin{multline*}
	\frac{\partial}{\partial x}f(s,x,u)= 
	\exp(-\zeta s)
	\frac{\partial}{\partial x}R(x)+
	\frac{\partial}{\partial x}g(s,x)\\\times
	[1+sb^2x\left(1+\mbox{$\frac{1}{2}$}s\sigma\right)]+
	sb^2(1+\mbox{$\frac{1}{2}$}s\sigma)g(s,x),
	\end{multline*}
	\[
	\frac{\partial}{\partial u}f(s,x,u)=-2cu\exp(-\zeta s),
	\]
	\begin{multline*}
	\frac{\partial^2}{\partial x^2}f(s,x,u)=
	\exp(-\zeta s)\frac{\partial^2}{\partial x^2}R(x)+
	\frac{\partial^2}{\partial x^2}g(s,x)
	[1+sb^2x\left(1+\mbox{$\frac{1}{2}$}s\sigma\right)]\\+
	sb^2(1+\mbox{$\frac{1}{2}$}s\sigma)
	\left[g(s,x)+\frac{\partial}{\partial x}g(s,x)\right],
	\end{multline*}
	and
	\[
	\frac{\partial^2}{\partial x\partial u}f(s,x,u)=0.
	\]
	The right hand side of Equation (\ref{o}) becomes zero and the Walrasian optimal strategy is $\phi_w^*(s,x)=0$.
	
	The corresponding Hamiltonian-Jacobi-Bellman Equation is
	\begin{multline}\label{hjb0}
	-\frac{\partial}{\partial s}V(s,x)-\mbox{$\frac{1}{2}$}\sigma x 
	\frac{\partial^2}{\partial x^2}V(s,x)\\=
	\max_{u\in U}
	\left\{\exp(-\zeta s)[R(x)-cu^2]+bx 
	\frac{\partial}{\partial x}V(s,x)\right\}.
	\end{multline}
	After solving for the right hand side of Equation (\ref{hjb0}) we 
	get $\phi_w^*(s,x)=0$. In this example we conclude that if the trend of the market dynamics does not depend on $u(s)$, there is no optimal strategy under both of quantum approach and Pontryagin maximum principle. 
\end{example}

Another important example considers problems involving European call options, which have been well studied in finance, and provide the basis for the Black-Scholes formula and further generalizations by Merton-Garman. In the generalized approach the stock volatility is stochastic and is derived by a parabolic partial differential equation \citep{baaquie1997,merton1973}. As constructing a Hamiltonian-Jacobi-Bellman equation becomes impossible in this case, methods of theoretical physics have been applied to get an optimal solution \citep{bouchaud1994}. For example, the Feynman-Kac lemma has been used in \cite{baaquie1997} and \cite{belal2007} to find a solution of a Merton-Garman-Hamiltonian type equations using the Dirac bracket method \cite{bergmann}. In Proposition \ref{p2} we use a path integral approach to a situation where the firm's objective is to maximize its portfolio subject to a Merton-Garman-Hamiltonian type stochastic volatility in an European call option with controls. Using the function $g$ as defined for Proposition \ref{p1}, the result given below provides an optimal investment strategy for this framework.

For this type of problem suppose that the firm has the objective 
of maximizing 
\[
\Pi_{\text{MG}}(u,t)=
\mathbb{E}\int_0^t\pi[x,H(s,K,V),V(s),u(s)]ds,
\]
where $u(s)$ is the strategy, and $H$ is the European call option 
price which is a function of the time $s$, the stock price of the 
security at time $s$ is represented by $K(s)$, 
and the volatility at time $s$ is represented by $V(s)$. 
It is assumed that the stock price and the volatility follow 
Langevin dynamics of the form 
\[
dK(s)=\mu_1[s,u(s)]K(s)ds+\sigma_1[s,u(s)]K(s)dB_1(s),
\]
and 
\[
dV(s)=\mu_2[s,u(s)]V(s)ds+\sigma_2[s,u(s)]V(s)dB_2(s),
\]
where $\mu_1[s,u(s)]$ is the expected return of the security, $\mu_2[s,u(s)]$ is the expected rate of increase in $V(s)$, and $B_1(s)$ and $B_2(s)$ are standard Brownian motion processes such that the correlation between $dB_1(r)$ and $dB_2(s)$ is zero unless $s=r$ for which case it equals a value $\gamma\in[-1,1]$. 

\begin{prop}[Merton-Garman Hamiltonian Type Equation]\label{p2}
	Suppose that a firm's objective portfolio is given by
	maximizing $\Pi_{\text{MG}}(u,t)$ with respect to the strategy $u\in U$. Let
	\begin{eqnarray}\label{m13.0}
	f(s,K,V,u) & = & \pi[s,H(s,K,V),V,u]+g(s,K,V)+
	\frac{\partial}{\partial s}g(s,K,V)\notag\\ & &+ 
	K\mu_1(s,u)\frac{\partial}{\partial K}g(s,K,V)+
	 V\mu_2(s,u)\frac{\partial}{\partial V}g(s,K,V)\notag\\ & &+
	\mbox{$\frac{1}{2}$}K^2\sigma_1^2(s,u)
	\frac{\partial^2}{\partial K^2}g(s,K,V)+
	K\rho\sigma_1^3(s,u)\times\notag \\
	& &
	\frac{\partial^2}{\partial K\partial V}g(s,K,V) 
	 +\mbox{$\frac{1}{2}$}V^2\sigma_2^2(s,u)
	\frac{\partial^2}{\partial V^2}g(s,K,V). 
	\end{eqnarray}
	An optimal Walrasian strategy is the functional solution of 
	\[
	-\left[
	\frac{\partial}{\partial u}f(s,K,V,u)
	\right]
	\Psi_s(K,V)=0
	\]
	where $\Psi_s(K,V)=\exp\{-s f(s,K,V,u)\}I(K,V)$
	is the transition wave function at time $s$ and states $K(s)$ and $V(s)$ with initial condition $\Psi_0(K,V)=I(K,V)$.
\end{prop}
Proposition \ref{p2} is the extension of the framework of \cite{baaquie1997} that accounts for the firm's portfolio and has drift and diffusion components that are functions of the feedback control system and considers an optimal Walrasian strategy.

Proposition \ref{p3} considers the case of the cooperative 
environment outlined in Definition \ref{paretodef}.

\begin{prop}[Cooperative Pareto Optimality]\label{p3}
	A cooperative Pareto optimal solution for firm $\rho$ where all the firms maximize the total dynamic profit $\overline{\Pi}_\text{P}(u,t)$ subject to 
	\begin{equation*}
	d\mathbf{x}(s)=\bm{\mu}[s,\mathbf{x}(s),\mathbf{u}(s)]ds+\bm{\sigma}[s,\mathbf{x}(s),\mathbf{u}(s)]d\mathbf{B}(s),
	\end{equation*}
	with initial condition $\mathbf{x}(0)=\mathbf{x}_0$
	is obtained by solving 
	\begin{equation}\label{sch3}
	-\frac{\partial f[s,\mathbf{x}(s),u(s)]}{\partial u_\rho}\ \Psi_s(\mathbf x)=0,
	\end{equation}
	with respect to $\rho^{th}$ firm's strategy, where $\Psi_s$ is the transition wave function defined as 
	\[
	\Psi_s(\mathbf{x})=\exp[-f(s,\mathbf{x},u)]\Psi_0(\mathbf{x})
	\]
	with initial condition $\Psi_0(\mathbf{x})$ and $f$ is defined as 
	\begin{multline*}
	f(s,\mathbf{x},u)=
	\sum_{\rho=1}^k\alpha_\rho\pi_\rho(s,\mathbf{x},u)+g(s,\mathbf{x})+
	\frac{\partial}{\partial s}g(s,\mathbf{x})\\+
	\bm{\mu}'(s,\mathbf{x},\mathbf{u})\mathcal{D}_{ \mathbf{x}}g(s,\mathbf{x})+
	\mbox{$\frac{1}{2}$}\bm{\sigma}'(s,\mathbf{x})\mathcal{H}_{\mathbf{x}}g(s,\mathbf{x})\bm{\sigma}(s,\mathbf{x}),
	\end{multline*}
	where $\mathcal{D}_\mathbf{x}$ is the gradient vector and $\mathcal{H}_\mathbf{x}$ is the Hessian matrix. 
\end{prop}

\begin{example}\label{ex5}
	Suppose that a firm under a Cooperative Pareto system has the objective function 
	\[ 
	\E\left\{\int_0^t\exp(-r s)\sum_{\rho=1}^k\a_\rho\left(px_\rho-cx_\rho u_\rho^2\right)ds\right\},
	\]
	where $r\in(0,1]$ is a constant discount rate over time interval $[0,t]$, $p>0$ is constant price, $\a_\rho$ is the weight corresponding to $\rho^{th}$ firm such that $\sum_{\rho=1}^k\a_\rho=1$, $x_\rho$ is $\rho^{th}$ firm's total output, $c$ is a positive constant marginal cost for each firm, and $u_\rho$ is the total expenditure on advertising of the $\rho^{th}$ firm. 
	Consider market dynamics
	\[
	d\mathbf{x}(s)=
	[\mathbf{x}'(s)\mathbf{a}\mathbf{x}(s)-\mathbf{u}(s)]ds+
	\mathbf{x}(s)\bm{\sigma'}d\mathbf{B}(s),
	\]
	where $\mathbf{x}$ and $\mathbf{u}$ both are $k$-dimensional vectors such that $x_\rho\in\mathcal X$ and $u_\rho\in\mathcal U^\rho\in U$, $\mathbf a$ is a $k\times k$-dimensional constant symmetric matrix, $\bm\sigma$ is an $m$-dimensional constant vector and $\mathbf B$ is an $m$-dimensional Brownian motion process. 
	For a given Lagrangian multiplier $\lambda^*$ assume $g(s,\mathbf x)=s\lambda^*[\mathbf x'\mathbf a\mathbf x-b]$. 
	Therefore,
	\begin{multline*}
	f(s,\mathbf x,\mathbf u)=\exp(-r s)\sum_{\rho=1}^k\a_\rho\big(px_\rho-cx_\rho u_\rho^2\big)+(1+s)\lambda^*[\mathbf x'\mathbf a\mathbf x-b]\\+2s\lambda^*[\mathbf x'\mathbf a'\mathbf x-\mathbf u']\mathbf x'\mathbf a+s\lambda^*\bm\sigma\mathbf x'\mathbf a\mathbf x\bm\sigma'.
	\end{multline*}
	Equation (\ref{sch3}) implies,
	\[
	\phi_{p\rho}^*(s,\mathbf x)=\frac{s\lambda^*\mathbf x'\mathbf a'}{c\a_\rho x_\rho \exp(-rs)},
	\]
	such that $c\a_\rho x_\rho \exp(-rs)\neq 0$.
\end{example}

\begin{example}\label{ex5.0}
	Consider a resource extraction problem of two players as discussed in the Section $7.2.1$ of \cite{yeung2006}. 
	Suppose, there are two players with objective function
	\begin{multline*}
	\max_{u_1,u_2}\E\int_0^t\exp(-rs)
	\left\{\left[\left(k_1u_1(s)\right)^{1/2}-
	\frac{c_1u_1(s)}{\mathbf{x}^{1/2}(s)}\right]\right.\\+\left.
	\a_1^0\left[\left(k_2u_2(s)\right)^{1/2}-
	\frac{c_2u_2(s)}{\mathbf x^{1/2}(s)}\right]\right\}ds,
	\end{multline*}
	subject to 
	\[
	d\mathbf x(s)=\left[a\mathbf x^{\frac{1}{2}}(s)-b\mathbf x(s)-u_1(s)-u_2(s)\right]ds+\bm\sigma\mathbf x'(s)\ d\mathbf B(s).
	\]
	In the above problem $u^\rho\in\mathcal U^\rho\in\mathcal U$ is the control strategy vector of player $\rho$ for $\rho\in\{1,2\}$, $a$ and $b$ are positive constant scalar, $\bm\sigma$ is an $m$-dimensional constant, $\a_1^0\in[0,\infty)$ is the optimal cooperative weight corresponding to player $2$  and $B(s)$ is am $m$-dimensional Brownian motion. Here $[k_\rho u_\rho(s)]^{\frac{1}{2}}$ is player $\rho$'s level of satisfaction from the consumption of the resource extracted at time $s$ and $c-\rho u_\rho(s)\mathbf x^{-\frac{1}{2}}(s)$ is the dissatisfaction level brought about by the cost extraction. Finally, $k_1,k_2,c_1,c_2$ are positive constant scalars.
	
	(i) Quantum approach: For a given fixed Lagrange multiplier $\lambda^*$ and a positive constant scalar $d$ assume $g(s,\mathbf x)=s\lambda^*\big[a\mathbf x^{\frac{1}{2}}(s)-b\mathbf x(s)-d\big]$, where $d$ takes care of the variability coming from $\sum_{q=1}^{k}u_q^*(s)+u_\rho(s)$. Hence, $\frac{\partial}{\partial s}g(s,\mathbf x)=\lambda^*\big[a\mathbf x^{\frac{1}{2}}(s)-b\mathbf x(s)-d\big]$, $\mathcal D_{\mathbf x}g(s,\mathbf x)=s\lambda^*\left[\frac{a}{2}\mathbf x^{-\frac{1}{2}}-b\right]$ and $\mathcal{H}_{\mathbf x}g(s,\mathbf x)\\=-s\lambda^*\frac{a}{4}\mathbf x^{-\frac{3}{2}}$. Therefore,
	\begin{multline*}
	f(s,\mathbf x, u_1,u_2)=\exp(-rs)\left\{\left[\left(k_1u_1\right)^{\frac{1}{2}}-\frac{c_1u_1}{\mathbf x^{\frac{1}{2}}}\right]+\a_1^0\left[\left(k_2u_2\right)^{\frac{1}{2}}-\frac{c_2u_2}{\mathbf x^{\frac{1}{2}}(s)}\right]\right\}\\+(1+s)\lambda^*\left[a\mathbf x^{\frac{1}{2}}-b\mathbf x-d\right]\\+s\lambda^*\left[a\mathbf x^{\frac{1}{2}'}-b\mathbf x'-u_1-u_2\right]\left(\mbox{$\frac{a}{2}$}\mathbf x^{-\frac{1}{2}}-b\right)-s\lambda^*\mbox{$\frac{a}{8}$}\mathbf x \bm\sigma'\mathbf x^{-\frac{3}{2}}\bm\sigma\mathbf x'.
	\end{multline*}
	Equation (\ref{sch3}) gives us the cooperative Pareto optimal strategy of two players as
	\begin{align}
	\phi_{p1}^*(s,\mathbf x)&=\mbox{$\frac{1}{4}$}k_1\left[\frac{\exp(-rs)}{c_1\mathbf x^{-\frac{1}{2}}\exp(-rs)+s\lambda^*(\frac{a}{2}\mathbf x^{\frac{1}{2}}-b)}\right]^2,\notag\\ \phi_{p2}^*(s,\mathbf x)&=\mbox{$\frac{1}{4}$}k_2\left[\frac{\a_1^0\exp(-rs)}{\a_1^0c_2\mathbf x^{-\frac{1}{2}}\exp(-rs)+s\lambda^*(\frac{a}{2}\mathbf x^{\frac{1}{2}}-b)}\right]^2,\notag
	\end{align}
	where $c_1\mathbf x^{-\frac{1}{2}}\exp(-rs)+s\lambda^*(\frac{a}{2}\mathbf x^{\frac{1}{2}}-b)\neq 0$ and $\a_1^0c_2\mathbf x^{-\frac{1}{2}}\exp(-rs)+s\lambda^*(\frac{a}{2}\mathbf x^{\frac{1}{2}}-b)\neq 0$.
	
	(ii) Pontryagin maximum principle: From Example $7.2.1$ in \cite{yeung2006} we get cooperative Pareto optimal strategies of two players as,
	\begin{align}
	\phi_{p1}^*(s,\mathbf x)&=\frac{k_1\mathbf x}{4\left[c_1+\exp(-rs)\mathbf x^{\frac{1}{2}}\ \mathcal D_{\mathbf x}W^{\a_1^0}(s,\mathbf x)\right]^2},\notag\\
	\phi_{p2}^*(s,\mathbf x)&=\frac{k_2\mathbf x}{4\left[c_2+\frac{1}{\a_1^0}\exp(-rs)\mathbf x^{\frac{1}{2}}\ \mathcal D_{\mathbf x}W^{\a_1^0}(s,\mathbf x)\right]^2},\notag
	\end{align}
	where for $s\in[0,t]$ the value function is 
	\[
	W^{\a_1^0}(s,\mathbf x)=\exp(-rs)\left[A^{\a_1^0}(s) \mathbf x^{\frac{1}{2}} +B^{\a_1^0}(s)\right],
	\]
	such that, $A^{\a_1^0}(s)$ and $B^{\a_1^0}(s)$ satisfy:
	\[
	\mbox{$\frac{\partial}{\partial s}A^{\a_1^0}(s)$}=\left[r+\mbox{$\frac{1}{8}$}\bm\sigma'\bm\sigma+\mbox{$\frac{1}{2}$}b\right]A^{\a_1^0}(s)-\mbox{$\frac{k_1}{4\left[c_1+\frac{1}{2}A^{\a_1^0}(s)\right]}$}-\mbox{$\frac{\a_1^0 k_2}{4\left[c_2+\frac{1}{2\a_1^0}A^{\a_1^0}(s)\right]}$}
	\]
	and,
	\[
	\mbox{$\frac{\partial}{\partial s}B^{\a_1^0}(s)$}=rB^{\a_1^0}(s)-\mbox{$\frac{1}{2}$}a A^{\a_1^0}(s).
	\]
\end{example}

Finally, we find optimal strategy of the $\rho^{th}$ firm using a non-cooperative feedback Nash equilibrium. We assume that a firm is rational in decision making and earns more profit at the cost of the profit of the other firms in the market. Hence, Firm $\rho$ seeks to maximize 
\[
\Pi_{\text{N}}(u,t)=\E\int_0^t\pi_\rho[s,\mathbf{x}(s),u_\rho(s),\mathbf{u}_{-\rho}^*(s)]ds
\]
with respect to the strategy $u_\rho$ where $u_{-\rho}^*(s)$ is the optimized strategies for firms other than the $\rho^{th}$ firm.

\begin{prop}\label{pr1}
	A non-cooperative Nash optimal solution for maximizing $\Pi_{\text{N}}(u,t)$
	subject to 
	\begin{equation*}
	d\mathbf{x}(s)=\bm{\mu}[s,\mathbf{x}(s),u_\rho(s),\mathbf{u}_{-\rho}^*(s)]ds+\bm{\sigma}[s,\mathbf{x}(s),u_\rho(s),\mathbf{u}_{-\rho}^*(s)]d\mathbf{B}(s),
	\end{equation*} 
	with initial condition $\mathbf{x}(0)=\mathbf{x}_0$ is the solution of
	\begin{equation}\label{na2.3}
	-\frac{\partial f^\rho[s,\mathbf{x}(s),u_\rho(s),\mathbf{u}_{-\rho}^*(s)]}{\partial u_\rho}\ \Psi_s(\mathbf{x})=0,
	\end{equation}
	where $\Psi_s$ is the transition wave function defined as 
	\[
	\Psi_s(\mathbf{x})=\exp[-f(s,\mathbf{x},u_\rho(s),\mathbf{u}_{-\rho}^*(s))]\Psi_0(\mathbf{x})
	\]
	with initial condition $\Psi_0(\mathbf{x})$ and
	\begin{eqnarray*}
		f^\rho[s,\mathbf{x},u_\rho(s),\mathbf{u}_{-\rho}^*(s)] & = & \pi_\rho[s,\mathbf{x}(s),u_\rho(s),\mathbf{u}_{-\rho}^*(s)]\\ & &+g^\rho[s,\mathbf{x}(s)]+\frac{\partial}{\partial s}g^\rho[s,\mathbf{x}(s)]\\ & &+\bm{\mu}'[s,\mathbf{x}(s),u_\rho(s),\mathbf{u}_{-\rho}^*(s)]\mathcal{D}_{\mathbf{x}}g^\rho[s,\mathbf{x}(s)]\\
		& & +\mbox{$\frac{1}{2}$}\bm{\sigma}'[s,\mathbf{x}(s),u_\rho(s),u_{-\rho}^*(s)]\mathcal{H}_{\mathbf{x}}g^\rho[s,\mathbf{x}(s)]\\ & &\times\bm{\sigma}'[s,\mathbf{x}(s),u_\rho(s),u_{-\rho}^*(s)].
	\end{eqnarray*}
\end{prop} 

\begin{example}\label{ex6}
	Consider an economy endowed with a renewable resource with $k\geq 2$ firms such as in section $2.6$  in \cite{yeung2006}. We can compare our Nash equilibrium strategy through quantum approach with traditional Pontryagin maximum principle in \cite{yeung2006}. Suppose, $\rho^{th}$ firm's resource extraction in time $s\in[0,t]$ is $u_\rho(s)$ for all $\rho=\{1,2,...,k\}$. Define $\mathbf{u}_{-\rho}^*=\sum_{q=1}^{k}u_{q}^*(s)$ where $\rho\neq q$ and $k$-dimensional vector $\mathbf{x}(s)$ is the size of the resource stock at time $s$ such that $\mathbf x(s)>\bm 0$. Under this construction $\rho^{th}$ firm's objective function is
	\[
	\E\left\{\int_0^t\exp(-rs)\left[\left(\sum_{q=1}^{k}u_q^*(s)+u_\rho\right)^{-\frac{1}{2}}u_\rho(s)-\frac{c}{\mathbf x^{\frac{1}{2}}(s)}u_\rho(s)\right]ds\right\},
	\]
	subject to the resource dynamics
	\[
	d\mathbf x(s)=\left[a\mathbf x^{\frac{1}{2}}(s)-b\mathbf x(s)-\sum_{q=1}^{k}u_q^*(s)-u_\rho(s)\right]ds+\bm\sigma\mathbf x'(s)\ d\mathbf B(s),
	\]
	where $cu_\rho(s)/[\mathbf x^{\frac{1}{2}}(s)]$ is $\rho^{th}$ firm's cost of resource extraction at time $s$, $\bm\sigma$ is am $m$-dimensional constant diffusion vector component and, vector $\mathbf B(s)$ is an $m$-dimensional Brownian motion. In this model assume $a,b$ and $c$ are the scalars. For a given fixed Lagrange multiplier $\lambda^*$ assume $g^\rho(s,\mathbf x)=s\lambda^*\left[a\mathbf x^{\frac{1}{2}}(s)-b\mathbf x(s)-d\right]$, where $d$ takes care of the variability coming from $\sum_{q=1}^{k}u_q^*(s)+u_\rho(s)$. Hence, $\frac{\partial}{\partial s}g^\rho(s,\mathbf x)=\lambda^*\left[a\mathbf x^{\frac{1}{2}}(s)-b\mathbf x(s)-d\right]$, $\mathcal D_{\mathbf x}g^\rho(s,\mathbf x)\\=s\lambda^*\left[\frac{a}{2}\mathbf x^{-\frac{1}{2}}-b\right]$ and $\mathcal{H}_{\mathbf x}g^\rho(s,\mathbf x)=-s\lambda^*\frac{a}{4}\mathbf x^{-\frac{3}{2}}$. Therefore,
	\begin{align}
	f^\rho(s,\mathbf x,u_\rho,u_{-\rho}^*)&=\exp(-rs)\left[\left(\sum_{q=1}^{k}u_q^*+u_\rho\right)^{-\frac{1}{2}}u_\rho-\frac{c}{\mathbf x^{\frac{1}{2}}}u_\rho\right]\notag\\&+(1+s)\lambda^*\left[a\mathbf x^{\frac{1}{2}}-b\mathbf x-d\right]\notag\\&+s\lambda^*\left[a\mathbf x^{\frac{1}{2}'}-b\mathbf x'-\sum_{q=1}^{k}u_q^*-u_\rho\right]\left(\mbox{$\frac{a}{2}$}\mathbf x^{-\frac{1}{2}}-b\right)\notag\\&-s\lambda^*\mbox{$\frac{a}{8}$}\mathbf x \bm\sigma'\mathbf x^{-\frac{3}{2}}\bm\sigma\mathbf x'.\notag
	\end{align}
	Finally, Equation (\ref{na2.3}) implies the feedback Nash Equilibrium as
	\begin{align}
	\phi_{NQ}^{\rho*}(s,\mathbf x)=2\left(\sum_{q=1}^{k}u_q^*\right)^{\frac{3}{2}}\left[\frac{c}{\mathbf x^{\frac{1}{2}}}+s\lambda^*\exp(rs)\left(\mbox{$\frac{a}{2}$}\mathbf x^{-\frac{1}{2}}-b\right)-\left(\sum_{q=1}^{k}u_q^*\right)^{-\frac{1}{2}}\right].\notag
	\end{align}
	From section $2.6$ of \cite{yeung2006} we know, the feedback Nash equilibrium from Pontryagin maximum principle is
	\begin{align}
	\phi_{NP}^{\rho*}(s,\mathbf x)&=\frac{\mathbf x(2k-1)^2}{2\left[\sum_{q=1}^k\left(c+\exp(rs)\mathcal D_{\mathbf x}V^q\mathbf x^{\frac{1}{2}}\right)\right]}\notag\\&\times \left\{\sum_{q=1}^k\left[c+\frac{\mathcal D_{\mathbf x}V^q\mathbf x^{\frac{1}{2}}}{\exp(-rs)}\right]-\left(k-\frac{3}{2}\right)\left[c+\frac{\mathcal D_{\mathbf x}V^\rho\mathbf x^{\frac{1}{2}}}{\exp(-rs)}\right]\right\},\notag
	\end{align}
	where $V^\rho$ and $V^q$ are the value function of firms $\rho$ and $q$ with their gradients $\mathcal D_{\mathbf x}V^\rho$ and $\mathcal D_{\mathbf x}V^q$ respectively. By Corollary $2.6.1$ in \cite{yeung2006} Hamiltonian-Jacobi-Bellman system has a solution
	\[
	V^\rho(s,\mathbf x)=\exp(-rs)\left[A(s)\mathbf x^{\frac{1}{2}}+B(s)\right],
	\]
	where $A(s)$ and $B(s)$ satisfies,
	\begin{align}
	\mbox{$\frac{\partial}{\partial s}$}A(s)&=\left[r+\mbox{$\frac{1}{8}$}\bm\sigma'\bm\sigma-\mbox{$\frac{b}{2}$}\right]A(s)-\frac{2k-1}{2k^2}\left[c+\mbox{$\frac{1}{2}$}A(s)\right]^{-1}\notag\\&\hspace{2cm}+\frac{c(2k-1)^2}{4k^3}\left[c+\mbox{$\frac{1}{2}$}A(s)\right]^{-2}+\frac{(2k-1)^2A(s)}{8k^2\left[c+\mbox{$\frac{1}{2}$}A(s)\right]^2},\notag\\\mbox{$\frac{\partial}{\partial s}$}B(s)&=rB(s)-\mbox{$\frac{1}{2}$}a A(s).\notag
	\end{align}
\end{example}

\section{Proofs}
\subsection{Proof of Proposition \ref{p1}}

The arguments here are based on the use of the quantum Lagrangian action function. Further details are given in the Appendix. Equation (\ref{mkt}) implies
$\Delta x(s)=x(s+ds)-x(s)=\mu[s,x(s),u(s)]ds+\sigma[s,x(s),u(s)]dB(s)$. Following \cite{chow1996} from Equation (\ref{w6}), the Euclidean action function is, 
\begin{multline*}
\mathcal{A}_{0,t}(x)=\int_0^t\E_s\{\pi[s,x(s),u(s)]ds+\lambda [\Delta x(s)\\-\mu[s,x(s),u(s)]ds-\sigma[s,x(s),u(s)]dB(s)]\}.
\end{multline*}
Let $\varepsilon>0$, and for a normalizing constant $L_\varepsilon>0$ from Lemma \ref{l1} in the Appendix, let
\begin{equation}\label{w16}
\Psi_{s,s+\varepsilon}(x)=\frac{1}{L_\varepsilon} \int_{\mathbb{R}}\exp[-\varepsilon \mathcal{A}_{s,s+\varepsilon}(x)]\Psi_s(x)dx(s),
\end{equation}
where $\Psi_s(x)$ is the value of the transition function at time $s$ and state $x(s)$ with the initial condition $\Psi_0(x)=\Psi_0$.

Fubini's Theorem implies that the action function on time interval $[s,s+\varepsilon]$ is
\[
\mathcal{A}_{s,s+\varepsilon}(x)=
\E_s\int_{s}^{s+\varepsilon}\pi[\nu,x(\nu),u(\nu)]d\nu+g[\nu+\Delta \nu,x(\nu)+\Delta x(\nu)].
\]
where 
\begin{multline*}
g[\nu+\Delta\nu,x(\nu)+\Delta x(\nu)]=\lambda[\Delta x(\nu)-\mu[\nu,x(\nu),u(\nu)]d\nu\\-\sigma[\nu,x(\nu),u(\nu)]dB(\nu)+o(1).
\end{multline*}
 This conditional expectation is valid when the strategy $u(\nu)$ is determined at time $\nu$ and the measure of firm's share $x(\nu)$ is known \citep{chow1996}. The evolution of a process takes place as if the action function is stationary. Therefore, the conditional expectation with respect to time only depends on the expectation of initial time point of this time interval.
 
 It\^o's Lemma implies,
 \begin{eqnarray*}
 	\varepsilon\mathcal{A}_{s,s+\varepsilon}(x) & = & 
 	\E_s\left\{\varepsilon\pi[s,x(s),u(s)]+\varepsilon g[s,x(s)]\right.\\ & & \left.+ \varepsilon\frac{\partial}{\partial s}g[s,x(s)]+ \varepsilon\mu[s,x(s),u(s)]\frac{\partial}{\partial x}g[s,x(s)]\right. \\
 	& & \left. +\varepsilon\sigma[s,x(s),u(s)]\frac{\partial}{\partial x}g[s,x(s)]dB(s)\right.\\ & & \left.+
 	\mbox{$\frac{1}{2}$}\varepsilon\sigma^2[s,x(s),u(s)]\frac{\partial^2}{\partial x^2}g[s,x(s)]+o(\varepsilon)\right\},
 \end{eqnarray*}
 and $[\Delta x(s)]^2\sim\varepsilon$ as $\varepsilon\rightarrow0$. \cite{feynman1948} uses an interpolation method to find an approximation of the area under the path in $[s,s+\varepsilon]$. Using a similar approximation, 
 \begin{eqnarray}
 \mathcal{A}_{s,s+\varepsilon}(x)& = & \pi[s,x(s),u(s)]+g[s,x(s)]+
 \frac{\partial}{\partial s}g[s,x(s)]\notag\\ & &+
 \mu[s,x(s),u(s)]
 \frac{\partial}{\partial x}g[s,x(s)] \notag \\
 & & +\mbox{$\frac{1}{2}$}
 \sigma^2[s,x(s),u(s)]
 \frac{\partial^2}{\partial x^2}g[s,x(s)]+o(1), \label{action}
 \end{eqnarray}
 where $\E_s[dB(s)]=0$ and $\E_s[o(\varepsilon)]/\varepsilon\ra 0$ as $\varepsilon\ra 0$.  Combining Equations (\ref{w16}) and (\ref{action}) yield
 \begin{eqnarray}\label{action9}
 \Psi_{s,s+\varepsilon}(x) & = & \frac{1}{L_\varepsilon}\int_{\mathbb{R}}
 \exp\left[-\varepsilon\left\{
 \pi[s,x(s),u(s)]+g[s,x(s)]+
 \frac{\partial}{\partial s}g[s,x(s)]\right.\right.\notag \\ & &\left.\left.+
 \mu[s,x(s),u(s)]\frac{\partial}{\partial x}g[s,x(s)]+
 \mbox{$\frac{1}{2}$}\sigma^2[s,x(s),u(s)]\frac{\partial^2}{\partial x^2}g[s,x(s)]\right\}\right]\notag\\ & & \times\Psi_s[x(s)]dx(s)+o(\varepsilon^{1/2}),
 \end{eqnarray}
 as $\varepsilon\ra 0$. Taking a first order Taylor series expansion on the left hand side of Equation (\ref{action9}) yields
 \begin{align}\label{action10}
 &\Psi_s^\tau(x)+\varepsilon  \frac{\partial \Psi_s^\tau(x)}{\partial s}+o(\varepsilon)\notag\\&=\frac{1}{L_\varepsilon}\int_{\mathbb{R}} \exp\biggr\{-\varepsilon \big[\pi[s,x(s),u(s)]\notag\\&+g[s,x(s)]+\frac{\partial}{\partial s}g[s,x(s)]+\frac{\partial}{\partial x}g[s,x(s)]\mu[s,x(s),u(s)]\notag\\&+\mbox{$\frac{1}{2}$}\sigma^2[s,x(s),u(s)]\frac{\partial^2}{\partial x^2}g[s,x(s)]\big]\biggr\}  \Psi_s[x(s)] dx(s)+o(\varepsilon^{1/2}).
 \end{align}
 For fixed $s$ and $\tau$ let $x(s)=x(\tau)+\xi$ and assume that for some $0<\eta<\infty$ we have $|\xi|\leq\sqrt{\frac{\eta\varepsilon}{x(s)}}$ so that $0<x(s)\leq\eta\varepsilon/\xi^2$. Furthermore, as our stochastic isoperimetric non-holonomic constraint follows Theorem \ref{thmf} along with Assumptions \ref{as1}, \ref{as2} in the Appendix, and $d\xi$ is a cylindrical measure where $\xi$ contributes significantly, $\Psi_\tau[x(\xi)] $ of Equation (\ref{action10}) can be expanded using a Taylor series of $\xi$ around $0$.
 Therefore,	
 \begin{align}
 &\Psi_s^\tau(x)+\varepsilon  \frac{\partial \Psi_s^\tau(x)}{\partial s}+o(\varepsilon)\notag\\&= \frac{1}{L_\varepsilon}\int_{\mathbb{R}} \left[\Psi_s^\tau(x)+\xi\frac{\partial \Psi_s^\tau(x)}{\partial x}+o(\varepsilon)\right] \exp\biggr\{-\varepsilon \big[\pi[s,x(\tau)+\xi,u(s)]\notag\\&+g[s,x(\tau)+\xi]+\frac{\partial}{\partial s}g[s,x(\tau)+\xi]+\frac{\partial}{\partial x}g[s,x(\tau)+\xi] \mu[s,x(\tau)+\xi,u(s)]\notag\\&+\mbox{$\frac{1}{2}$}\sigma^2[s,x(\tau)+\xi,u(s)] \frac{\partial^2}{\partial x^2}g[s,x(\tau)+\xi]\big]\biggr\} d\xi+o(\varepsilon^{1/2}).\notag
 \end{align}
 
 Let
 \begin{eqnarray*}
 	f[s,\xi,u(s)]&=&\pi[s,x(\tau)+\xi,u(s)]+g[s,x(\tau)+\xi]+\frac{\partial}{\partial s}g[s,x(\tau)+\xi]\\ & & +\frac{\partial}{\partial x}g[s,x(\tau)+\xi] \mu[s,x(\tau)+\xi,u(s)]\\ & &+\mbox{$\frac{1}{2}$}\sigma^2[s,x(\tau)+\xi,u(s)]\ \frac{\partial^2}{\partial x^2}g[s,x(\tau)+\xi]+o(1),
 \end{eqnarray*}
 so that
 \begin{multline*}
 \Psi_s^\tau(x)+\varepsilon  \frac{\partial \Psi_s^\tau(x)}{\partial s}+o(\varepsilon)=\Psi_s^\tau(x) \frac{1}{L_\epsilon} \int_{\mathbb{R}} \exp\big\{-\varepsilon f[s,\xi,u(s)]\} d\xi\\+\frac{\partial \Psi_s^\tau(x)}{\partial x}\frac{1}{L_\varepsilon}\int_{\mathbb{R}} \xi \exp \big\{-\varepsilon f[s,\xi,u(s)]\big\} d\xi+o(\varepsilon^{1/2}).
 \end{multline*}
 where
 \begin{multline*}
 f[s,\xi,u(s)]=f[s,x(\tau),u(s)]+\frac{\partial}{\partial x}f[s,x(\tau),u(s)][\xi-x(\tau)]\\+\mbox{$\frac{1}{2}$}\frac{\partial^2}{\partial x^2}f[s,x(\tau),u(s)][\xi-x(\tau)]^2+o(\varepsilon),
 \end{multline*}
 where $\varepsilon\ra 0$ and $\Delta x\ra 0$.
 
 Define $m=\xi-x(\tau)$ so that $ d\xi=dm$, then standard integration techniques can be used to show that
 \begin{multline*}
 \int_{\mathbb{R}} \exp\left\{-\varepsilon f[s,\xi,u(s)]\right\} d\xi\notag=\exp\{-\varepsilon f[s,x(\tau),u(s)]\}\times\\\int_{\mathbb{R}} \exp\biggr\{-\varepsilon \biggr[\frac{\partial}{\partial x}f[s,x(\tau),u(s)]m+\mbox{$\frac{1}{2}$}\frac{\partial^2}{\partial x^2}f[s,x(\tau),u(s)]m^2\biggr]\biggr\} dm.
 \end{multline*}
 Then 
 \begin{multline*}
 \Psi_s^\tau(x) \frac{1}{L_\varepsilon} \int_{\mathbb{R}} \exp\big\{-\varepsilon f[s,\xi,u(s)]\} d\xi\\=\Psi_s^\tau(x) \frac{1}{L_\varepsilon}  \sqrt{\frac{\pi}{\varepsilon a}}\exp\left\{\varepsilon \left[\frac{b^2}{4a^2}-f[s,x(\tau),u(s)]\right]\right\}, 
 \end{multline*}
 where $a=\frac{1}{2} \frac{\partial^2}{\partial x^2}f[s,x(\tau),u(s)]$ and $b=\frac{\partial}{\partial x}f[s,x(\tau),u(s)]$.
 
 Similarly, it can be shown that
 \begin{align}
 &\frac{\partial\Psi_s^\tau(x)}{\partial x} \frac{1}{L_\varepsilon}\int_{\mathbb{R}} \xi \exp\left[-\varepsilon f[s,\xi,u(s)]\right] d\xi\notag\\&=\frac{\partial\Psi_s^\tau(x)}{\partial x} \frac{1}{L_\varepsilon} \exp\left\{\varepsilon \left[\frac{b^2}{4a^2}-f[s,x(\tau),u(s)]\right]\right\} \biggr[x(\tau)-\frac{b}{2a}\biggr]\sqrt{\frac{\pi}{\varepsilon a}}.\notag
 \end{align}
 Therefore
 \begin{multline*}
 \Psi_s^\tau(x)+\varepsilon  \frac{\partial \Psi_s^\tau(x)}{\partial s}+o(\varepsilon)=\Psi_s^\tau(x) \frac{1}{L_\varepsilon} \sqrt{\frac{\pi}{\varepsilon a}}\exp\left\{\varepsilon \left[\frac{b^2}{4a^2}-f[s,x(\tau),u(s)]\right]\right\}\\+\frac{\partial\Psi_s^\tau(x)}{\partial x} \frac{1}{L_\varepsilon} \sqrt{\frac{\pi}{\varepsilon a}} \exp\left\{\varepsilon \left[\frac{b^2}{4a^2}-f[s,x(\tau),u(s)]\right]\right\} \biggr[x(\tau)-\frac{b}{2a}\biggr]+o(\varepsilon^{1/2}).
 \end{multline*}
 Assuming $L_\varepsilon=\sqrt{\frac{\pi}{\varepsilon a}}>0$ and after expanding exponential function up to the first order we get,
 \begin{multline*}
 \Psi_s^\tau(x)+\varepsilon \frac{\partial \Psi_s^\tau(x)}{\partial s}+o(\varepsilon)=\left\{1+\varepsilon \left[\frac{b^2}{4a^2}-f[s,x(\tau),u(s)]\right]\right\}\\\times \biggr\{\Psi_s^\tau(x)+\left[x(\tau)-\frac{b}{2a}\right]\frac{\partial\Psi_s^\tau(x)}{\partial x}+o(\varepsilon^{1/2})\biggr\}.
 \end{multline*}
 
 The term $b/(2a)$ is the ratio of the first derivative to the second derivative with respect to $x$ of $f$. As $f$ is in a Schwartz space, the derivatives of $f$ are rapidly falling and they satisfy Assumptions \ref{as1} and \ref{as2}, and therefore it is reasonable to we assume, $0<|b|\leq\eta\varepsilon$ and $0<|a|\leq\mbox{$\frac{1}{2}$}(1-\xi^{-2})^{-1}$. Hence, using $x(s)-x(\tau)=\xi$ we get,
 \begin{align*}
 x(\tau)-\frac{b}{2a}=x(s)-\frac{b}{2a}.
 \end{align*}
 and therefore
 \begin{align*}
 \bigg|x(s)-\frac{b}{2a}\bigg|\leq\eta\varepsilon.
 \end{align*}
 
 Therefore, letting $\varepsilon\ra 0$, the Wick rotated Schr\"odinger type equation is,
 \begin{align}\label{action25.4}
 \frac{\partial \Psi_s^\tau(x)}{\partial s}&=\left[\frac{b^2}{4a^2}-f[s,x(\tau),u(s)]\right]\Psi_s^\tau(x).
 \end{align}
 If we differentiate Equation (\ref{action25.4}) with respect to $u$, then the solution of the new equation will be a Walrasian optimal strategy in the stochastic case. That is,
 \begin{multline}\label{w18}
 \left[\frac{2\frac{\partial}{\partial x}f(s,x,u)}{\frac{\partial^2}{\partial x^2}f(s,x,u)}\left(\frac{\frac{\partial^2}{\partial x^2}f(s,x,u)\frac{\partial}{\partial x\partial u}f(s,x,u)-\frac{\partial}{\partial x}f(s,x,u)\frac{\partial^3}{\partial u\partial x^2}f(s,x,u)}{\left[\frac{\partial^2}{\partial x^2}f(s,x,u)\right]^2}\right)\right.\\\left.-\frac{\partial}{\partial u}f(s,x,u)\right]\Psi_s^\tau(x)=0.
 \end{multline}
 Therefore, an optimal Walrasian strategy is found by setting Equation (\ref{w18}) equal to zero obtains,
 \begin{align*}
 \frac{\partial}{\partial u}f(s,x,u) \left[\frac{\partial^2}{\partial x^2}f(s,x,u)\right]^2=2\frac{\partial}{\partial x}f(s,x,u) \frac{\partial^2}{\partial x\partial u}f(s,x,u).
 \end{align*}
 A unique solution to Equation (\ref{action25.4}) can be found using a Fourier transformation, as $\Psi_s(x)=I(x) \exp[sv(x,u)]$, which can be verified by direct differentiation. $\square$

 \subsection{Proof of Proposition \ref{p2}}
 
 Euclidean action function is,
 \begin{multline*}
 \mathcal{A}_{0,t}(K,V)= \int_0^t \E_s\bigg\{\pi\bigg[s,H[s,K(s),V(s)],u(s)\bigg]\ ds+\lambda_1 \big[K(s+ds)-K(s)\\-\mu_1[s,u(s)] K(s) ds-\sigma_1[s,u(s)] K(s) dB_1(s)\big] \\ +\lambda_2\big[V(s+ds)-V(s)-\mu_2[s,u(s)]\ V(s)\ ds-\sigma_2[s,u(s)] V(s) dB_2(s)\big]\bigg\}.
 \end{multline*}
 Following arguments similar to those used to prove Proposition \ref{p1}, define $\Delta s=\varepsilon>0$, and for $L_\varepsilon>0$ Lemma \ref{l1} in the Appendix implies, 
 \begin{align}\label{m3}
 \Psi_{s,s+\varepsilon}(K,V)&= \frac{1}{L_\varepsilon} \int_{\mathbb{R}} \exp\biggr\{-\varepsilon  \mathcal{A}_{s,s+\varepsilon}(K,V)\biggr\} \Psi_s(K,V) dK(s)\times dV(s),
 \end{align}
 as $\varepsilon\ra 0$ where $\Psi_s(K,V)$ is the wave function at time $s$ and states $K(s)$ and $V(s)$ respectively with initial condition $\Psi_0(K,V)=\Psi_0$.
 
 The action function in $[s,\tau]$ where $\tau=s+\varepsilon$ with the Lagrangian is,
 \begin{multline*}
 \mathcal{A}_{s,\tau}(K,V)= \int_s^\tau\ \E_s\bigg\{\pi\bigg[\nu,H[\nu,K(\nu),V(\nu)],V(\nu),u(\nu)\bigg] d\nu\\+\lambda_1 \big[K(\nu+d\nu)-K(\nu)-\mu_1[\nu,u(\nu)] K(\nu) d\nu-\sigma_1[\nu,u(\nu)]\ K(\nu) dB_1(\nu)\big] \notag\\+\lambda_2 \big[V(\nu+d\nu)-V(\nu)-\mu_2[\nu,u(\nu)] V(\nu) d\nu-\sigma_2[\nu,u(\nu)] V(\nu) dB_2(\nu)\big]\bigg\},
 \end{multline*}
 with initial conditions $K(0)=K_0$ and $V(0)=V_0$, where $\lambda_1$ and $\lambda_2$ are two Lagrangian multipliers corresponding to the two constraints. The conditional expectation is valid when the strategy $u(\nu)$ is determined at time $\nu$, and hence only depends on the initial time point of this time interval.
 Let $\Delta K(\nu)=K(\nu+d\nu)-K(\nu)$ and, $\Delta V(\nu)=V(\nu+d\nu)-V(\nu)$, then Fubini's Theorem implies,
 \begin{align}\label{m5}
 \mathcal{A}_{s,\tau}(K,V)&=\int_s^\tau\ \E_s\bigg\{\pi\bigg[\nu,H[\nu,K(\nu),V(\nu)],V(\nu),u(\nu)\bigg] d\nu\notag \\&+\lambda_1 \big[\Delta K(\nu)-\mu_1[\nu,u(\nu)] K(\nu) d\nu-\sigma_1[\nu,u(\nu)] K(\nu)\ dB_1(\nu)\big]\notag\\ & +\lambda_2\big[\Delta V(\nu)-\mu_2[\nu,u(\nu)]\ V(\nu)\ d\nu-\sigma_2[\nu,u(\nu)]\ V(\nu)\ dB_2(\nu)\big]\bigg\}.
 \end{align}
 Because $K(\nu)$ and $V(\nu)$ are It\^o processes, Theorem 4.1.2 of \cite{oksendal2003} implies that there exists a function\\ $g[\nu,K(\nu),V(\nu)]\in C^2([0,\infty)\times\mathbb{R}\times \mathbb{R})$ that satisfies Theorem \ref{thmf} in the Appendix, Assumptions \ref{as1} and \ref{as2}, such that $Y(\nu)=g[\nu,K(\nu),V(\nu)]$ where $Y(\nu)$ is an It\^o process. If we assume 
 \begin{multline*}
 g[\nu+\Delta \nu,K(\nu)+\Delta K(\nu),V(\nu)+\Delta V(\nu)]\\=\lambda_1 \big[\Delta K(\nu)-\mu_1[\nu,u(\nu)] K(\nu) d\nu-\sigma_1[\nu,u(\nu)] K(\nu) dB_1(\nu)\\+\lambda_2  \big[\Delta V(\nu)-\mu_2[\nu,u(\nu)]\ V(\nu) d\nu-\sigma_2[\nu,u(\nu)]\ V(\nu)\ dB_2(\nu)+o(1),
 \end{multline*} 
 Equation (\ref{m5}) becomes,
 \begin{align}\label{m6}
 \mathcal{A}_{s,\tau}(K,V)&=\E_s\ \bigg\{ \int_{s}^\tau \pi\bigg[\nu,H[\nu,K(\nu),V(\nu)],V(\nu),u(\nu)\bigg]\ d\nu\notag\\&\hspace{1cm}+g[\nu+\Delta \nu,K(\nu)+\Delta K(\nu),V(\nu)+\Delta V(\nu)]\bigg\}.
 \end{align}
 It\^o's Lemma and Equation (\ref{m6}) of \cite{baaquie1997} imply
 \begin{align}
 \mathcal{A}_{s,\tau}(K,V) &=\pi\bigg[s,H[s,K(s),V(s)],V(s),u(s)\bigg]+g[s,K(s),V(s)]\notag\\ &+\frac{\partial}{\partial s}g[s,K(s),V(s)]+\frac{\partial}{\partial S}g[s,K(s),V(s)] \mu_1[s,u(s)] K(s)\notag\\&+\frac{\partial}{\partial V}g[s,K(s),V(s)] \mu_2[s,u(s)] V(s) \notag\\&+\mbox{$\frac{1}{2}$}\bigg[\sigma_1^2[s,u(s)] K^2(s) \frac{\partial^2}{\partial K^2}g[s,K(s),V(s)]\notag\\&+2\rho\sigma_1^3[s,u(s)] K(s) \frac{\partial^2}{\partial K\partial V}g[s,K(s),V(s)]\notag\\&+\sigma_2^2[s,u(s)] V^2(s) \frac{\partial^2}{\partial V^2}g[s,K(s),V(s)]\bigg]+o(1),\notag
 \end{align}
 where we have used the fact that $[\Delta K(s)]^2=\Delta V(s)]^2=\varepsilon$, and $\E_s[\Delta B_1(s)]=\E_s[\Delta B_2(s)]$, as $\varepsilon\ra 0$ with initial conditions $K_0$ and $V_0$. Using Equation (\ref{m3}), the transition wave function in $[s,\tau]$ becomes,
 \begin{align}
 &\Psi_{s,\tau}(K,V)\notag\\&=\frac{1}{L_\varepsilon}\int_{\mathbb{R}^2} \exp\bigg\{-\varepsilon \bigg[\pi\bigg[s,H[s,K(s),V(s)],V(s),u(s)\bigg]+g[s,K(s),V(s)]\notag\\&+\frac{\partial}{\partial s}g[s,K(s),V(s)]+\frac{\partial}{\partial K}g[s,K(s),V(s)] \mu_1[s,u(s)] K(s)\notag\\&+\frac{\partial}{\partial V}g[s,S(s),V(s)] \mu_2[s,u(s)] V(s) +\mbox{$\frac{1}{2}$}\bigg[\sigma_1^2[s,u(s)] K^2(s) \frac{\partial^2}{\partial K^2}g[s,K(s),V(s)]\notag\\&+2\rho\sigma_1^3[s,u(s)] K(s) \frac{\partial^2}{\partial K\partial V}g[s,K(s),V(s)]\notag\\&+\sigma_2^2[s,u(s)] V^2(s) \frac{\partial^2}{\partial V^2}g[s,K(s),V(s)]\bigg]\bigg] \bigg\} \Psi_s(K,V) dK(s) dV(s)+o(\varepsilon^{1/2}),\notag
 \end{align}
 as $\varepsilon\ra 0$. 
 
 Therefore,
 \begin{align}
 &\Psi_s^\tau(K,V)+\varepsilon \frac{\partial \Psi_s^\tau(K,V)}{\partial\notag s}+o(\varepsilon)\notag\\&=\frac{1}{L_\varepsilon}\int_{\mathbb{R}^2} \exp\bigg\{-\varepsilon \bigg[\pi\bigg[s,H[s,K(s),V(s)],V(s),u(s)\bigg]+g[s,K(s),V(s)]\notag\\&+\frac{\partial}{\partial s}g[s,K(s),V(s)]+\frac{\partial}{\partial K}g[s,K(s),V(s)] \mu_1[s,u(s)]\ K(s)\notag\\&+\frac{\partial}{\partial V}g[s,K(s),V(s)] \mu_2[s,u(s)] V(s) +\mbox{$\frac{1}{2}$}\bigg[\sigma_1^2[s,u(s)] K^2(s)\notag\\ &\times \frac{\partial^2}{\partial K^2}g[s,K(s),V(s)]+2\rho\sigma_1^3[s,u(s)] K(s) \frac{\partial^2}{\partial K\partial V}g[s,K(s),V(s)]\notag\\&+\sigma_2^2[s,u(s)] V^2(s) \frac{\partial^2}{\partial V^2}g[s,K(s),V(s)]\bigg]\bigg] \bigg\}\Psi_s(K,V) dK(s) dV(s)+o(\varepsilon^{1/2}),\notag
 \end{align}
 as $\varepsilon\ra 0$.
 
 \bigskip
 
 For fixed $s$ and $\tau$ suppose that $K(s)=K(\tau)+\xi_1$, and $V(s)=V(\tau)+\xi_2$. For positive numbers $\eta_1<\infty$ and $\eta_2<\infty$  assume that $|\xi_1|\leq\sqrt{\frac{\eta_1\epsilon}{K(s)}}$ and $|\xi_2|\leq\sqrt{\frac{\eta_2\epsilon}{V(s)}}$. Here, security and volatility are $K(s)\leq\eta_1\epsilon/\xi_1^2$ and $V(s)\leq\eta_2\epsilon/\xi_2^2$, respectively. Furthermore, Theorem \ref{thmf} and Assumptions \ref{as1}, \ref{as2} in the Appendix imply
 \begin{align}
 &\Psi_s^\tau(K,V)+\varepsilon \frac{\partial \Psi_s^\tau(K,V)}{\partial\notag s}+o(\varepsilon)\notag\\&=\frac{1}{L_\varepsilon}\int_{\mathbb{R}^2} \left[\Psi_s^\tau(K,V)+\xi_1\frac{\partial \Psi_s^\tau(K,V)}{\partial K}+\xi_2\frac{\partial \Psi_s^\tau(K,V)}{\partial V}+o(\varepsilon)\right]\notag\\& \exp\bigg\{-\varepsilon \bigg[\pi\bigg[s,H[s,K(\tau)+\xi_1,V(\tau)+\xi_2],V(\tau)+\xi_2,u(s)\bigg]\notag\\&+g[s,K(\tau)+\xi_1,V(\tau)+\xi_2]+\frac{\partial}{\partial s}g[s,K(\tau)+\xi_1,V(\tau)+\xi_2]\notag\\&+g_K[s,K(\tau)+\xi_1,V(\tau)+\xi_2] \mu_1[s,u(s)] (K(\tau)+\xi_1)\notag\\&+\frac{\partial}{\partial V}g[s,K(\tau)+\xi_1,V(\tau)+\xi_2] \mu_2[s,u(s)] (V(\tau)+\xi_2) \notag\\&+\mbox{$\frac{1}{2}$}\bigg[\sigma_1^2[s,u(s)](K(\tau)+\xi_1)^2 \frac{\partial^2}{\partial K^2}g[s,K(\tau)+\xi_1,V(\tau)+\xi_2]\notag\\&+2\rho\sigma_1^3[s,u(s)] (K(\tau)+\xi_1) \frac{\partial^2}{\partial K\partial V}g[s,K(\tau)+\xi_1,V(\tau)+\xi_2]\notag\\&+\sigma_2^2[s,u(s)] (V(\tau)+\xi_2)^2 \frac{\partial^2}{\partial V^2}g[s,K(\tau)+\xi_1,V(\tau)+\xi_2]\bigg]\bigg] \bigg\}\notag\\& \Psi_\tau[K(\xi_1),V(\xi_2)]\ d\xi_1 d\xi_2+o(\varepsilon^{1/2}),\notag
 \end{align}
 as $\varepsilon\ra 0$.
 
 Define $f[s,\xi_1,\xi_2,u(s)]$ as in Equation (\ref{m13.0}), then 
 \begin{align}
 &\Psi_s^\tau(K,V)+\varepsilon \frac{\partial \Psi_s^\tau(K,V)}{\partial\notag s}+o(\varepsilon)\notag\\&=\frac{1}{L_\varepsilon} \Psi_s^\tau(K,V) \int_{\mathbb{R}^2} \exp \big\{-\varepsilon f[s,\xi_1,\xi_2,u(s)]\big\}d\xi_1 d\xi_2\notag\\&+\frac{1}{L_\varepsilon} \frac{\partial \Psi_s^\tau(K,V)}{\partial K} \int_{\mathbb{R}^2} \xi_1 \exp \big\{-\varepsilon f[s,\xi_1,\xi_2,u(s)]\big\} d\xi_1 d\xi_2\notag\\&+\frac{1}{L_\varepsilon} \frac{\partial \Psi_s^\tau(K,V)}{\partial V} \int_{\mathbb{R}^2} \xi_2 \exp \big\{-\varepsilon f[s,\xi_1,\xi_2,u(s)]\big\} d\xi_1 d\xi_2+o(\varepsilon^{1/2}).\notag
 \end{align}
 Assume that $f$ is a $C^2$ function, then
 \begin{align}
 & f[s,\xi_1,\xi_2,u(s)]\notag\\&=f[s,K(\tau),V(\tau),u(s)]+[\xi_1-K(\tau)] \frac{\partial}{\partial K}f[s,K(\tau),V(\tau),u(s)]\notag\\&+[\xi_2-V(\tau)] \frac{\partial}{\partial V}f[s,K(\tau),V(\tau),u(s)]\notag\\&+\mbox{$\frac{1}{2}$}\bigg[[\xi_1-K(\tau)]^2\frac{\partial^2}{\partial K^2}f[s,K(\tau),V(\tau),u(s)]\notag\\&+2 [\xi_1-K(\tau)] [\xi_2-V(\tau)] \frac{\partial^2}{\partial K\partial V}g[s,K(\tau),V(\tau),u(s)]\notag\\&+[\xi_2-V(\tau)]^2 \frac{\partial^2}{\partial V^2}g[s,K(\tau),V(\tau),u(s)]\bigg]+o(\varepsilon),\notag
 \end{align}
 as $\varepsilon\ra 0$ and $\Delta u\ra 0$.
 Define $m_1=\xi_1-K(\tau)$ and $m_2=\xi_2-V(\tau)$ so that $d\xi_1=dm_1$ and $d\xi_2=dm$ respectively so that
 \begin{align}\label{m16}
 & \int_{\mathbb{R}^2} \exp \big\{-\varepsilon f[s,\xi_1,\xi_2,u(s)]\big\} d\xi_1 d\xi_2\notag\\&=\int_{\mathbb{R}^2} \exp \biggr\{-\varepsilon \bigg[f[s,K(\tau),V(\tau),u(s)]+m_1 \frac{\partial}{\partial K}f[s,K(\tau),V(\tau),u(s)]\notag\\&\hspace{.5cm}+m_2\frac{\partial}{\partial V}f[s,K(\tau),V(\tau),u(s)]+\mbox{$\frac{1}{2}$} m_1^2 \frac{\partial^2}{\partial K^2}g[s,K(\tau),V(\tau),u(s)]\notag\\&\hspace{1cm}+m_1 m_2 \frac{\partial^2}{\partial K\partial V}f[s,K(\tau),V(\tau),u(s)]\notag\\&\hspace{2cm}+\mbox{$\frac{1}{2}$} m_2^2 \frac{\partial^2}{\partial V^2}f[s,K(\tau),V(\tau),u(s)]\bigg]\biggr\} dm_1 dm_2.
 \end{align}
 Let
 \[
 \Theta=\begin{bmatrix}
 \mbox{$\frac{1}{2}$} \frac{\partial^2}{\partial K^2}f[s,K(\tau),V(\tau),u(s)] & \mbox{$\frac{1}{2}$} \frac{\partial^2}{\partial K\partial V}g[s,K(\tau),V(\tau),u(s)] \\ \mbox{$\frac{1}{2}$} \frac{\partial^2}{\partial K\partial V}f[s,K(\tau),V(\tau),u(s)]& \mbox{$\frac{1}{2}$} \frac{\partial^2}{\partial V^2}g[s,K(\tau),V(\tau),u(s)]
 \end{bmatrix},
 \]
 and
 \[
 m=\begin{bmatrix}
 m_1\\ m_2
 \end{bmatrix},
 \] 
 and 
 \[ 
 -v_1=\begin{bmatrix}
 \frac{\partial}{\partial K}f[s,K(\tau),V(\tau),u(s)]\\ \frac{\partial}{\partial V}f[s,K(\tau),V(\tau),u(s)]
 \end{bmatrix},
 \]
 where we assume that $\Theta$ is positive definite, then the integrand in Equation (\ref{m16}) becomes a shifted Gaussian integral, 
 \begin{align}
 & \int_{\mathbb{R}^2} \exp \bigg\{-\varepsilon \left(f- v_1^T\ m+m^T \Theta m\right)\bigg\}\ dm\notag\\&=\exp\left(-\varepsilon f\right) \int_{\mathbb{R}^2} \exp \bigg\{(\varepsilon v_1^T) m-m^T (\varepsilon \Theta) m\bigg\} dm=\frac{\pi}{\sqrt{\varepsilon |\Theta|}} \exp\left[\frac{\varepsilon}{4}v_1^T \Theta^{-1} v_1-\varepsilon f\right],\notag
 \end{align}
 where $v_1^T$ and $m^T$ are the transposes of vectors $v_1$ and $m$ respectively. Therefore,
 \begin{align}\label{m19}
 &\frac{1}{L_\varepsilon} \Psi_s^\tau(K,V) \int_{\mathbb{R}^2} \exp \big\{-\epsilon f[s,\xi_1,\xi_2,u(s)]\big\} d\xi_1 d\xi_2\notag\\&\hspace{1.5cm}=\frac{1}{L_\varepsilon} \Psi_s^\tau(K,V) \frac{\pi}{\sqrt{\varepsilon |\Theta|}}\exp\left[\frac{\varepsilon}{4} v_1^T \Theta^{-1} v_1-\varepsilon f\right],
 \end{align}
 such that inverse matrix $\Theta^{-1}>0$ exists. Similarly,
 \begin{align}\label{m21}
 & \frac{1}{L_\varepsilon} \frac{\partial \Psi_s^\tau(K,V)}{\partial K} \int_{\mathbb{R}^2} \xi_1 \exp \big\{-\varepsilon f[s,\xi_1,\xi_2,u(s)]\big\} d\xi_1 d\xi_2\notag\\&=\frac{1}{L_\varepsilon} \frac{\partial \Psi_s^\tau(K,V)}{\partial K} \frac{\pi}{\sqrt{\varepsilon |\Theta|}} \left(\mbox{$\frac{1}{2}$} \Theta^{-1}+K\right) \exp\left[\frac{\varepsilon}{4}\ v_1^T\ \Theta^{-1}\ v_1-\varepsilon f\right],
 \end{align}
 and 
 \begin{align}\label{m22}
 & \frac{1}{L_\varepsilon} \frac{\partial \Psi_s^\tau(K,V)}{\partial V} \int_{\mathbb{R}^2} \xi_2 \exp \big\{-\varepsilon f[s,\xi_1,\xi_2,u(s)]\big\} d\xi_1\times d\xi_2\notag\\&=\frac{1}{L_\varepsilon}\frac{\partial \Psi_s^\tau(K,V)}{\partial V} \frac{\pi}{\sqrt{\varepsilon |\Theta|}} \left(\mbox{$\frac{1}{2}$} \Theta^{-1}+V\right) \exp\left[\frac{\varepsilon}{4} v_1^T \Theta^{-1}\ v_1-\varepsilon f\right].
 \end{align}
 Equations (\ref{m19}), (\ref{m21}) and (\ref{m22}) imply that the Wick rotated Schr\"odinger type equation is,
 \begin{multline*}
 \Psi_s^\tau(K,V)+\varepsilon \frac{\partial \Psi_s^\tau(K,V)}{\partial\notag s}+o(\varepsilon)\\=\frac{1}{L_\varepsilon} \frac{\pi}{\sqrt{\varepsilon |\Theta|}} \exp\left[\frac{\varepsilon}{4} v_1^T \Theta^{-1} v_1-\varepsilon f\right] \bigg[\Psi_s^\tau(K,V)+\left(\mbox{$\frac{1}{2}$} \Theta^{-1}+K\right) \frac{\partial \Psi_s^\tau(K,V)}{\partial K}\\+\left(\mbox{$\frac{1}{2}$} \Theta^{-1}+V\right) \frac{\partial \Psi_s^\tau(K,V)}{\partial V} \bigg]+o(\varepsilon^{1/2}),
 \end{multline*}
 as $\varepsilon\ra 0$.
 
 Assuming $L_\varepsilon=\pi/\sqrt{\varepsilon\ |\Theta|}>0$, 
 \begin{align}
 &\Psi_s^\tau(K,V)+\varepsilon \frac{\partial \Psi_s^\tau(K,V)}{\partial\notag s}+o(\varepsilon)\notag\\&=\left[1+\varepsilon\left(\frac{1}{4} v_1^T \Theta^{-1} v_1- f\right)\right] \bigg[\Psi_s^\tau(K,V)+\left(\mbox{$\frac{1}{2}$} \Theta^{-1}+K\right) \frac{\partial \Psi_s^\tau(K,V)}{\partial K}\notag\\&\hspace{2cm}+\left(\mbox{$\frac{1}{2}$} \Theta^{-1}+V\right) \frac{\partial \Psi_s^\tau(K,V)}{\partial V} \bigg]+o(\varepsilon^{1/2}).\notag
 \end{align}
 
 As $K(s)\leq \eta_1\varepsilon/\xi_1^2$, assume $|\Theta^{-1}|\leq 2\eta_1\varepsilon(1-\xi_1^{-2})$ such that $|(2\Theta)^{-1}+K|\leq\eta_1\varepsilon$. For $V(s)\leq \eta_2\varepsilon/\xi_2^2$ we assume $|\Theta^{-1}|\leq 2\eta_2\varepsilon(1-\xi_2^{-2})$ such that $|(2\Theta)^{-1}+V|\leq\eta_2\varepsilon$. Therefore, $|\Theta^{-1}|\leq 2\varepsilon\min\left\{\eta_1(1-\xi_1^{-2}),\eta_2(1-\xi_2^{-2})\right\}$ such that, $|(2\Theta)^{-1}+K|\ra 0$ and $|(2\Theta)^{-1}+V|\ra 0$. Hence
 \begin{align*}
 \Psi_s^\tau(K,V)+\varepsilon \frac{\partial \Psi_s^\tau(K,V)}{\partial s}+o(\varepsilon)&=(1-\varepsilon f) \Psi_s^\tau(K,V)+o(\varepsilon^{1/2}). 
 \end{align*}
 Therefore, the Wick rotated Schr\"odinger type Equation is,
 \begin{equation*}
 \frac{\partial \Psi_s^\tau(K,V)}{\partial s}=-f[s,\xi_1,\xi_2,u(s)]\ \Psi_s^\tau(K,V).
 \end{equation*}
 Therefore, the solution of
 \begin{equation}\label{m27}
 -\frac{\partial f[s,\xi_1,\xi_2,u(s)]}{\partial u}\ \Psi_s^\tau(K,V)=0,
 \end{equation}
 is a Walrasian optimal strategy, which has the form \[\Psi_s(K,V)=\exp\left\{-s f[s,\xi_1,\xi_2,u(s)]\right\} I(K,V).\] As the transition function $\Psi_s^\tau(K,V)$ is the solution to Equation (\ref{m27}), the result follows. $\square$
 
\subsection{Proof of Proposition \ref{p3}}

The Euclidean action function for firm $\rho$ under Pareto optimality in real time $[0,t]$ is, 
\begin{align}
\mathcal{A}_{0,t}(x)&=  \int_0^t \E_s\bigg\{\sum_{\rho=1}^k \a_\rho\pi_\rho[s,x(s),u(s)]\ ds+\lambda \big[x(s+ds)-x(s)\notag\\&\hspace{2cm}-\mu[s,x(s),u(s)] ds-\sigma[s,x(s),u(s)]\ dB(s)\big] \bigg\}.\notag
\end{align} 

Following the arguments for the proof of Proposition \ref{p1}, we have
\begin{align}
\mathcal{A}_{s,\tau}(x)&= \E_s \bigg\{ \int_{s}^{\tau} \sum_{\rho=1}^k \a_\rho \pi_\rho[\nu,x(\nu),u(\nu)] d\nu+\lambda(\nu+d\nu)\ \big[\Delta x(\nu)\notag\\&\hspace{2cm}-\mu[\nu,x(\nu),u(\nu)] d\nu-\sigma[\nu,x(\nu),u(\nu)]\ dB(\nu)\big] \bigg\},\notag
\end{align}
where $\tau=s+\varepsilon$.

As $x(\nu)$ is an It\^o process then from Theorem 4.1.2 of \cite{oksendal2003} we know there exists a $p$-dimensional vector valued function $g[\nu,x(\nu)]\in C^2([0,\infty)\times\mathbb{R}^n)$ that satisfies Theorem \ref{thmf} in the Appendix, Assumptions \ref{as1} and \ref{as2}, and $Y(\nu)=g[\nu,x(\nu)]$ where $Y(\nu)$ is an It\^o process. Assume 
\begin{multline*}
g[\nu+\Delta \nu,x(\nu)+\Delta x(\nu)]=\lambda \big[\Delta x(\nu)-\mu[\nu,x(\nu),u(\nu)]\ d\nu\\-\sigma[\nu,x(\nu),u(\nu)]\ dB(\nu)+o(1),
\end{multline*} 
as $\varepsilon\ra 0$, then the generalized It\^o's Lemma implies,
\begin{align}
& \mathcal{A}_{s,\tau}(x) \varepsilon\notag\\&= \E_s \bigg\{\sum_{\rho=1}^k \a_\rho \pi_\rho[s,x(s),u(s)] \varepsilon+g[s,x(s)] \varepsilon+\frac{\partial}{\partial s}g[s,x(s)] \varepsilon\notag\\ &+\sum_{i=1}^{nk}\frac{\partial}{\partial x_i}g[s,x(s)] \mu[s,x(s),u(s)] \varepsilon+\sum_{i=1}^{nk}\frac{\partial}{\partial x_i}g[s,x(s)] \sigma[s,x(s),u(s)] \varepsilon \Delta B(s)\notag\\&+\mbox{$\frac{1}{2}$} \sum_{i=1}^{nk}\sum_{j=1}^{nk} \sigma^{ij}[s,x(s),u(s)]\frac{\partial^2}{\partial x_i x_j}g[s,x(s)] \varepsilon+o(\varepsilon)\bigg\},\notag
\end{align}
where $\sigma^{ij}[s,x(s),u(s)]$ represents $\{i,j\}^{th}$ component of the variance-covariance matrix, and we used the conditions $\Delta B_i \Delta B_j=\delta^{ij} \varepsilon$, $\Delta B_i \varepsilon=\varepsilon \Delta B_i=0$, and $\Delta x_i(s) \Delta x_j(s)=\varepsilon$, where $\delta^{ij}$ is the Kronecker delta function.  Hence
\begin{multline*}
\mathcal{A}_{s,\tau}(x)= \bigg[\sum_{\rho=1}^k \a_\rho\pi_\rho[s,x(s),u(s)]+g[s,x(s)]+\frac{\partial}{\partial s}g[s,x(s)]\\+\sum_{i=1}^{nk}\frac{\partial}{\partial x_i}g[s,x(s)] \mu[s,x(s),u(s)]\\+\mbox{$\frac{1}{2}$} \sum_{i=1}^{nk}\sum_{j=1}^{nk} \sigma^{ij}[s,x(s),u(s)]\frac{\partial^2}{\partial x_i x_j}g[s,x(s)] +o(1)\bigg],
\end{multline*}
where $\E_s[\Delta B(s)]=0$ and $\E_s[o(\varepsilon)]/\varepsilon\ra 0$ as $\varepsilon\ra 0$ with the vector of initial conditions $x_{0_{nk\times 1}}$. Expanding $\Psi_{s,\tau}(x)$ yields,
\begin{multline*}
\Psi_s^\tau(x)+\varepsilon  \frac{\partial \Psi_s^\tau(x)}{\partial s}+o(\varepsilon)\\=\frac{1}{L_\varepsilon} \int_{\mathbb{R}^{n\times k}} \exp\biggr\{-\varepsilon  \bigg[\sum_{\rho=1}^k \a_\rho\pi_\rho[s,x(s),u(s)]+g[s,x(s)]\\+\frac{\partial}{\partial s}g[s,x(s)]+\sum_{i=1}^{nk}\frac{\partial}{\partial x_i}g[s,x(s)] \mu[s,x(s),u(s)]\\+\mbox{$\frac{1}{2}$} \sum_{i=1}^{nk}\sum_{j=1}^{nk} \sigma^{ij}[s,x(s),u(s)] \frac{\partial^2}{\partial x_i x_j}g[s,x(s)]\bigg]\biggr\} \Psi_s(x) dx(s)+o(\varepsilon^{1/2}).
\end{multline*}
Let $x(s)_{nk\times 1}=x(\tau)_{nk\times 1}+\xi_{nk\times 1}$ and assume $||\xi||\leq\eta\epsilon [x^T(s)]^{-1}$ for some $\eta>0$. Following previous arguments imply
\begin{align}
&\Psi_s^\tau(x)+\varepsilon  \frac{\partial \Psi_s^\tau(x)}{\partial s}+o(\varepsilon)\notag\\&= \frac{1}{L_\varepsilon}\int_{\mathbb{R}^{n\times k}} \left[\Psi_s^\tau(x)+\xi\frac{\partial \Psi_s^\tau(x)}{\partial x}+o(\varepsilon)\right] \exp\biggr\{-\varepsilon\bigg[\sum_{\rho=1}^k \a_\rho\pi_\rho[s,x(\tau)+\xi,u(s)]\notag\\&+g[s,x(\tau)+\xi]+\frac{\partial}{\partial s}g[s,x(\tau)+\xi]+\sum_{i=1}^{nk}\frac{\partial}{\partial x_i}g[s,x(\tau)+\xi]\ \mu[s,x(\tau)+\xi,u(s)]\notag\\&+\mbox{$\frac{1}{2}$}\sum_{i=1}^{nk}\sum_{j=1}^{nk} \sigma^{ij}[s,x(\tau)+\xi,u(s)] \frac{\partial^2}{\partial x_i \partial x_j}g[s,x(\tau)+\xi]\bigg]\biggr\} d\xi+o(\varepsilon^{1/2}).\notag
\end{align}
Let 
\begin{multline*}
f[s,\xi,u(s)]=\sum_{\rho=1}^k \a_\rho\pi_\rho[s,x(\tau)+\xi,u(s)]+g[s,x(\tau)+\xi]\\+\frac{\partial}{\partial s}g[s,x(\tau)+\xi]+\sum_{i=1}^{nk}\frac{\partial}{\partial x_i}g[s,x(\tau)+\xi]\ \mu[s,x(\tau)+\xi,u(s)]\\+\mbox{$\frac{1}{2}$}\sum_{i=1}^{nk}\sum_{j=1}^{nk}\ \sigma^{ij}\ [s,x(\tau)+\xi,u(s)]\frac{\partial^2}{\partial x_i\partial x_j}g[s,x(\tau)+\xi],
\end{multline*}
then
\begin{align}\label{p17}
&\Psi_s^\tau(x)+\varepsilon  \frac{\partial \Psi_s^\tau(x)}{\partial s}+o(\varepsilon)\notag\\&= \frac{1}{L_\varepsilon} \Psi_s^\tau(x) \int_{\mathbb{R}^{n\times k}} \exp\bigg\{-\varepsilon f[s,\xi,u(s)]\bigg\}  d\xi\notag\\&\hspace{1cm}+\frac{1}{L_\varepsilon} \frac{\partial \Psi_s^\tau(x)}{\partial x} \int_{\mathbb{R}^{n\times k}} \xi\exp\bigg\{-\varepsilon f[s,\xi,u(s)]\bigg\} d\xi+o(\varepsilon^{1/2}).
\end{align}
Expanding $f[s,\xi,u(s)]$ and defining $m_{nk\times 1}=\xi_{nk\times 1}-x(\tau)_{nk\times 1}$ so that $ d\xi\\=dm$, first integral on the right hand side of Equation (\ref{p17}) becomes, 
\begin{align}
&\int_{\mathbb{R}^{n\times k}} \exp\bigg\{-\varepsilon f[s,\xi,u(s)]\bigg\} d\xi\notag\\&=\exp\bigg\{-\varepsilon f[s,x(\tau),u(s)]\bigg\} \int_{\mathbb{R}^{n\times k}} \exp\biggr\{-\varepsilon \biggr[\sum_{i=1}^{nk} \frac{\partial}{\partial x_i}f[s,x(\tau),u(s)]\ m_i\notag\\ &\hspace{1cm}+\mbox{$\frac{1}{2}$} \sum_{i=1}^{nk}\sum_{j=1}^{nk} \frac{\partial^2}{\partial x_i \partial x_j}f[s,x(\tau),u(s)]\ m_im_j\biggr]\biggr\} dm+ o(\varepsilon).\notag
\end{align}
Assume there exists a symmetric, positive definite and non-singular Hessian matrix $\theta_{nk\times nk}$ and a vector ${w}_{nk\times 1}$ such that,
\begin{align*}
&\int_{\mathbb{R}^{n\times k}}\ \exp\bigg\{-\varepsilon f[s,\xi,u(s)]\bigg\} d\xi\\ &=\sqrt{\frac{(2\pi)^{nk}}{\varepsilon |\theta|}} \exp\bigg\{-\varepsilon f[s,x(\tau),u(s)]+\frac{\varepsilon}{2} w^T\theta^{-1}w\bigg\},
\end{align*}
where, 
\[
\theta=\begin{bmatrix}
\frac{\partial^2}{\partial x_1\partial x_1}f & \frac{\partial^2}{\partial x_1\partial x_2}f & \dots & \frac{\partial^2}{\partial x_1\partial x_{nk}}f\\ \frac{\partial^2}{\partial x_2\partial x_1}f & \frac{\partial^2}{\partial x_2\partial x_2}f & \dots & \frac{\partial^2}{\partial x_2\partial x_{nk}}f\\ \vdots & \vdots & \ddots &\vdots\\\frac{\partial^2}{\partial x_{nk}\partial x_1}f & \frac{\partial^2}{\partial x_{nk}\partial x_2}f & \dots & \frac{\partial^2}{\partial x_{nk}\partial x_{nk}}f
\end{bmatrix},
\]
and 
\[
w[s,x(\tau),u(s)]=\begin{bmatrix}
-\frac{\partial}{\partial x_1}f[s,x(\tau),u(s)]\\-\frac{\partial}{\partial x_2}f[s,x(\tau),u(s)]\\\vdots\\-\frac{\partial}{\partial x_{nk}}f[s,x(\tau),u(s)]
\end{bmatrix}.
\]
The second integral on the right hand side of Equation (\ref{p17}) becomes,
\begin{multline*}
\int_{\mathbb{R}^{n\times k}} \xi \exp\bigg\{-\varepsilon f[s,\xi,u(s)]\bigg\} d\xi\\=\sqrt{\frac{(2\pi)^{nk}}{\varepsilon |\theta|}} \exp\bigg\{-\varepsilon f[s,x(\tau),u(s)]+\frac{\varepsilon}{2} w^T\theta^{-1}w\bigg\} \bigg[x(\tau)+\mbox{$\frac{1}{2}$}\ (\theta^{-1}\ w)\bigg].
\end{multline*}
So that
\begin{align}
&\Psi_s^\tau(x)+\varepsilon  \frac{\partial \Psi_s^\tau(x)}{\partial s}+o(\varepsilon)\notag\\&=\frac{1}{L_\varepsilon} \sqrt{\frac{(2\pi)^{nk}}{\varepsilon |\theta|}}\exp\bigg\{-\varepsilon f[s,x(\tau),u(s)]+\mbox{$\frac{1}{2}$}\varepsilon w^T\theta^{-1}w\bigg\}\notag\\&\times\bigg[\Psi_s^\tau(x)+\left[x(\tau)+\mbox{$\frac{1}{2}$} (\theta^{-1}\ w)\right]\frac{\partial \Psi_s^\tau(x)}{\partial x}\bigg]+o(\varepsilon^{1/2}).\notag
\end{align}
Assume $L_\varepsilon=\sqrt{(2\pi)^{nk}/(\varepsilon |\theta|)}>0$, then
\begin{align}
&\Psi_s^\tau(x)+\varepsilon  \frac{\partial \Psi_s^\tau(x)}{\partial s}+o(\varepsilon)\notag\\&=\left\{1-\varepsilon f[s,x(\tau),u(s)]+\mbox{$\frac{1}{2}$}\varepsilon w^T\theta^{-1}w\right\}\notag\\ &\hspace{1cm}\times \bigg[\Psi_s^\tau(x)+\left[x(\tau)+\mbox{$\frac{1}{2}$} (\theta^{-1} w)\right]\frac{\partial \Psi_s^\tau(x)}{\partial x}\bigg]+o(\varepsilon^{1/2}).\notag
\end{align}
For any finite positive number $\eta$ we know $x(\tau)\leq\eta\varepsilon|\xi^T|^{-1}$, and there exists $|\theta^{-1}w|\leq 2 \eta\varepsilon|1-\xi^T|^{-1}$ such that for $\varepsilon\ra 0$ we have, $\big|x(\tau)+\mbox{$\frac{1}{2}$} (\theta^{-1} w)\big|\leq\eta\varepsilon$ and hence
\begin{align*}
\frac{\partial \Psi_s^\tau(x)}{\partial s}&=\left\{- f[s,x(\tau),u(s)]+\mbox{$\frac{1}{2}$} w^T\theta^{-1}w\right\} \Psi_s^\tau(x). 
\end{align*}
Taking $\varepsilon\ra 0$, the Wick rotated Schr\"odinger type equation is 
\begin{align*}
\frac{\partial \Psi_s^\tau(x)}{\partial s}&=- f[s,x(\tau),u(s)]\ \Psi_s^\tau(x),
\end{align*}
with the Wheeler-Di Witt type equation,
\begin{align*}
-\frac{\partial f[s,x(\tau),u(s)]}{\partial u_\rho}\ \Psi_s^\tau(x)=0,
\end{align*}
whose solution with respect to $u_\rho$ gives $\rho^{th}$ firm's cooperative Pareto Optimal strategy $\phi_p^{\rho*}[s,x^*(s)]$. $\square$

\subsection{Proof of Proposition \ref{pr1}}

Following \cite{chow1996} the Euclidean action function of firm $\rho$ in $[0,t]$ is, 
\begin{align}
\mathcal{A}_{0,t}^\rho(x)&=  \int_0^t \E_s\bigg\{ \pi_\rho[s,x(s),u_\rho(s),u_{-\rho}^*(s)] ds+\lambda_\rho \big[x(s+ds)-x(s)\notag\\&-\mu[s,x(s),u_\rho(s),u_{-\rho}^*(s)] ds-\sigma[s,x(s),u_\rho(s),u_{-\rho}^*(s)] dB(s)\big] \bigg\}.\notag
\end{align}
 Let $\Delta s=\varepsilon>0$, and for $L_\varepsilon>0$ from Lemma \ref{l1} in the Appendix, the transition wave function of firm $\rho$ is 
\begin{align}\label{na4}
\Psi_{s,s+\varepsilon}^\rho(x)&= \frac{1}{L_\varepsilon} \int_{\mathbb{R}^n} \exp\biggr\{-\varepsilon  \mathcal{A}_{s,s+\varepsilon}^\rho(x)\biggr\} \Psi_s^\rho(x) dx(s),
\end{align}
for a time interval $[s,s+\varepsilon]$ where $\varepsilon\ra 0$ and $\Psi_s^\rho(x)$ is the value of firm $\rho$'s transition function at time $s$ and states $x(s)$ with initial conditions $\Psi_0^\rho(x)=\Psi_0^\rho$. In Equation (\ref{na4}), $\mathbb{R}^n$ represents $n$-dimensional strategy space of firm $\rho$. Let $\Delta x(\nu)=x(\nu+d\nu)-x(\nu)$ then the Euclidean action function of firm $\rho$ is,
\begin{align}\label{na6}
\mathcal{A}_{s,\tau}^\rho(x)&= \E_s\ \bigg\{ \int_{s}^{\tau} \pi_i[\nu,x(\nu),u_\rho(\nu),u_{-\rho}^*(\nu)]\ d\nu+\lambda_i \big[\Delta x(\nu)\notag\\&\hspace{2cm}-\mu[\nu,x(\nu),u(\nu)] d\nu-\sigma[\nu,x(\nu),u_\rho(\nu),u_{-\rho}^*(\nu)] dB(\nu)\big] \bigg\}.
\end{align}
By Theorem 4.1.2 of \cite{oksendal2003} we know there exists a $p$-dimensional vector valued function $g^\rho[\nu,x(\nu)]\in C^2([0,\infty)\times\mathbb{R}^n)$ that satisfies Theorem \ref{thmf} in the Appendix, Assumptions \ref{as1} and \ref{as2}, and $Y^\rho(\nu)=g^\rho[\nu,x(\nu)]$ where $Y^\rho(\nu)$ is firm $\rho$'s It\^o process. If we assume 
\begin{multline*}
g^\rho[\nu+\Delta \nu,x(\nu)+\Delta x(\nu)]=\lambda_\rho \big[\Delta x(\nu)-\mu[\nu,x(\nu),u_\rho(\nu),u_{-\rho}^*(\nu)] d\nu\\-\sigma[\nu,x(\nu),u_\rho(\nu),u_{-\rho}^*(\nu)] dB(\nu)+o(1),
\end{multline*} 
Equation (\ref{na6}) becomes,
\begin{align}
\mathcal{A}_{s,\tau}^\rho(x)&=\E_s\ \bigg\{ \int_{s}^{\tau}\ \pi_\rho[\nu,x(\nu),u_\rho(\nu),u_{-\rho}^*(\nu)]\ d\nu+g^\rho[\nu+\Delta \nu,x(\nu)+\Delta x(\nu)]\bigg\}.\notag
\end{align}
Generalized It\^o's Lemma implies
\begin{multline*}
\mathcal{A}_{s,\tau}^\rho(x)= \bigg[\pi_\rho[s,x(s),u_\rho(s),u_{-\rho}^*(s)]+g^\rho[s,x(s)]\\+\frac{\partial}{\partial s}g^\rho[s,x(s)]+\sum_{i=1}^n\frac{\partial}{\partial x_i}g^\rho[s,x(s)] \mu[s,x(s),u_\rho(s),u_{-\rho}^*(s)]\\+\mbox{$\frac{1}{2}$} \sum_{i=1}^n\sum_{j=1}^n \sigma^{ij}\left[s,x(s),u_\rho(s),u_{-\rho}^*(s)\right] \frac{\partial^2}{\partial x_i\partial x_j}g^\rho[s,x(s)] +o(1)\bigg],
\end{multline*}
where $\E_s[\Delta B(s)]=0$ and $\E_s[o(\varepsilon)]/\varepsilon\ra 0$ as $\varepsilon\ra 0$ with the vector of initial conditions $x_0^\rho$, where\\ $\sigma^{ij}[s,x(s),u_\rho(s),u_{-\rho}^*(s)]$ represents $\{i,j\}^{th}$ component of the variance-covariance matrix, and $\Delta B_i\Delta B_j=\delta^{ij} \varepsilon$, $\Delta B_i \varepsilon=\varepsilon \Delta B_i=0$, and $\Delta x_i(s) \Delta x_j(s)=\varepsilon$. A Taylor series expansion of the vector valued transition function $\Psi_{s,\tau}^\rho$ implies
\begin{multline*}
\Psi_s^{\tau, \rho}(x)+\varepsilon  \frac{\partial \Psi_s^{\tau,\rho}(x)}{\partial s}+o(\varepsilon)\\=\frac{1}{L_\varepsilon} \int_{\mathbb{R}^n} \exp\biggr\{-\varepsilon  \bigg[\pi_\rho[s,x(s),u_\rho(s),u_{-\rho}^*(s)]\\+g^\rho[s,x(s)] \mu[s,x(s),u_\rho(s),u_{-\rho}^*(s)]\\+\mbox{$\frac{1}{2}$} \sum_{i=1}^n\sum_{j=1}^{n} \sigma^{ij}[s,x(s),u_\rho(s),u_{-\rho}^*(s)]\\\times \frac{\partial^2}{\partial x_i\partial x_j}g^\rho[s,x(s)]\bigg]\biggr\}\Psi_s^\rho(x) dx(s)+o(\varepsilon^{1/2}),
\end{multline*}
as $\varepsilon\ra 0$. Let $x(s)_{n\times 1}=x(\tau)_{n\times 1}+\xi_{n\times 1}$. There exists a positive number $\eta<\infty$ such that, $||\xi||\leq\eta\varepsilon [x^T(s)]^{-1}$, and $[x^T(s)]^{-1}$ exists and not equal to zero. Following our previous arguments
\begin{align}\label{na13}
&\Psi_s^{\tau,\rho}(x)+\varepsilon \frac{\partial \Psi_s^{\tau,\rho}(x)}{\partial s}+o(\varepsilon)\notag\\&= \frac{1}{L_\varepsilon}\int_{\mathbb{R}^n} \left[\Psi_s^{\tau,\rho}(x)+\xi\frac{\partial \Psi_s^{\tau,\rho}(x)}{\partial x}+o(\varepsilon)\right] \notag\\ &\times\exp\biggr\{-\varepsilon \bigg[\pi_\rho[s,x(\tau)+\xi,u_\rho(s),u_{-\rho}^*(s)]+g^\rho[s,x(\tau)+\xi]\notag\\&+\frac{\partial}{\partial s}g^\rho[s,x(\tau)+\xi]+\sum_{i=1}^n \frac{\partial}{\partial x_i}g^\rho[s,x(\tau)+\xi]\ \mu[s,x(\tau)+\xi,u_\rho(s),u_{-\rho}^*(s)]\notag\\&\hspace{1cm}+\mbox{$\frac{1}{2}$}\sum_{i=1}^n\sum_{j=1}^{n} \sigma^{ij} [s,x(\tau)+\xi,u_\rho(s),u_{-\rho}^*(s)]\notag\\ &\hspace{2cm}\times \frac{\partial^2}{\partial x_i\partial x_j}g^\rho[s,x(\tau)+\xi]\bigg]\biggr\} d\xi+o(\epsilon^{1/2}).
\end{align}
Let
\begin{multline*}
f^\rho[s,\xi,u_\rho(s),u_{-\rho}^*(s)]=\pi_\rho[s,x(\tau)+\xi,u_\rho(s),u_{-\rho}^*(s)]+g^\rho[s,x(\tau)+\xi]\\+\frac{\partial}{\partial s}g^\rho[s,x(\tau)+\xi]+\sum_{i=1}^n\frac{\partial}{\partial x_i}g^\rho[s,x(\tau)+\xi]\ \mu[s,x(\tau)+\xi,u_\rho(s),u_{-\rho}^*(s)]\\+\mbox{$\frac{1}{2}$}\sum_{i=1}^n\sum_{j=1}^{n}\ \sigma^{ij} [s,x(\tau)+\xi,u_\rho(s),u_{-\rho}^*(s)] \frac{\partial^2}{\partial x_i\partial x_j}g^\rho[s,x(\tau)+\xi]. 
\end{multline*}
Equation (\ref{na13}) then
\begin{align}
&\Psi_s^{\tau,\rho}(x)+\varepsilon  \frac{\partial \Psi_s^{\tau,\rho}(x)}{\partial s}+o(\varepsilon)\notag\\&= \frac{1}{L_\varepsilon} \Psi_s^{\tau,\rho}(x) \int_{\mathbb{R}^n} \exp\bigg\{-\varepsilon f^\rho[s,\xi,u_\rho(s),u_{-\rho}^*(s)]\bigg\}  d\xi\notag\\&+\frac{1}{L_\varepsilon} \frac{\partial \Psi_s^{\tau,\rho}(x)}{\partial x} \int_{\mathbb{R}^n} \xi\exp\bigg\{-\varepsilon f^\rho[s,\xi,u_\rho(s),u_{-\rho}^*(s)]\bigg\} d\xi+o(\varepsilon^{1/2}),\notag
\end{align}
where
\begin{multline*}
f^\rho[s,\xi,u_\rho(s),u_{-\rho}^*(s)]=f^\rho[s,x(\tau),u_\rho(s),u_{-\rho}^*(s)]\\+\sum_{i=1}^n \frac{\partial}{\partial x_i}f^\rho[s,x(\tau),u_\rho(s),u_{-\rho}^*(s)] [\xi_i-x_i(\tau)]\\+\mbox{$\frac{1}{2}$} \sum_{i=1}^n\sum_{j=1}^{n} \frac{\partial^2}{\partial x_i\partial x_j}f^\rho[s,x(\tau),u_\rho(s),u_{-\rho}^*(s)][\xi_i-x_i(\tau)] [\xi_j-x_j(\tau)] +o(\varepsilon),
\end{multline*}
as $\varepsilon\ra 0$. Define $m_{n\times 1}=\xi_{n\times 1}-x(\tau)_{n\times 1}$ so that $ d\xi=dm$, then
\begin{align}
&\int_{\mathbb{R}^n} \exp\bigg\{-\varepsilon f^\rho[s,\xi,u_\rho(s),u_{-\rho}^*(s)]\bigg\} d\xi\notag\\&=\sqrt{\frac{(2\pi)^n}{\varepsilon |\tilde{\theta}|}} \exp\bigg\{-\varepsilon f^\rho[s,x(\tau),u_\rho(s),u_{-\rho}^*(s)]+\mbox{$\frac{1}{2}$}\varepsilon \widehat{w}^T\tilde{\theta}^{-1}\widehat{w}\bigg\},\notag
\end{align} 
where  we assume $\tilde{\theta}_{n\times n}$ the symmetric, positive definite and non-singular Hessian matrix
\[
\tilde{\theta}=\begin{bmatrix}
\frac{\partial^2}{\partial x_1\partial x_1}f^\rho &  \frac{\partial^2}{\partial x_1\partial x_2}f^\rho& \dots & \frac{\partial^2}{\partial x_1\partial x_n}f^\rho\\ \frac{\partial^2}{\partial x_2\partial x_1}f^\rho & \frac{\partial^2}{\partial x_2\partial x_2}f^\rho& \dots & \frac{\partial^2}{\partial x_2\partial x_n}f^\rho\\ \vdots &\vdots&\ddots & \vdots\\\frac{\partial^2}{\partial x_n\partial x_1}f^\rho & \frac{\partial^2}{\partial x_n\partial x_2}f^\rho&\dots& \frac{\partial^2}{\partial x_n\partial x_n}f^\rho
\end{bmatrix},
\]
and 
\[
\widehat{w}[s,x(\tau),u_\rho(s),u_{-\rho}^*(s)]=\begin{bmatrix}
-\frac{\partial}{\partial x_1}f^\rho[s,x(\tau),u_\rho(s),u_{-\rho}^*(s)]\\-\frac{\partial}{\partial x_2}f^\rho[s,x(\tau),u_\rho(s),u_{-\rho}^*(s)]\\\vdots\\-\frac{\partial}{\partial x_n}f^\rho[s,x(\tau),u_\rho(s),u_{-\rho}^*(s)]
\end{bmatrix}.
\]
Similarly,
\begin{multline*}
\int_{\mathbb{R}^n} \xi \exp\bigg\{-\varepsilon f^\rho[s,\xi,u_\rho(s),u_{-\rho}^*(s)]\bigg\} d\xi\\=\sqrt{\frac{(2\pi)^n}{\varepsilon |\tilde{\theta}|}} \exp\bigg\{-\varepsilon f^\rho[s,x(\tau),u_\rho(s),u_{-\rho}^*(s)]+\mbox{$\frac{1}{2}$}\varepsilon \widehat{w}^T\tilde{\theta}^{-1}\widehat{w}\bigg\} \bigg[x(\tau)+\mbox{$\frac{1}{2}$} (\theta^{-1} \widehat{w})\bigg],
\end{multline*}
and hence
\begin{multline*}
\Psi_s^{\tau,\rho}(x)+\varepsilon \frac{\partial \Psi_s^{\tau,\rho}(x)}{\partial s}+o(\varepsilon)\\=\frac{1}{L_\varepsilon} \sqrt{\frac{(2\pi)^n}{\varepsilon |\tilde{\theta}|}} \exp\bigg\{-\varepsilon f^\rho[s,x(\tau),u_\rho(s),u_{-\rho}^*(s)]+\mbox{$\frac{1}{2}$}\varepsilon \widehat{w}^T\tilde{\theta}^{-1}\widehat{w}\bigg\}\\\times\bigg[\Psi_s^{\tau,\rho}(x)+\left[x(\tau)+\mbox{$\frac{1}{2}$} (\tilde{\theta}^{-1} \widehat{w})\right]\frac{\partial \Psi_s^{\tau,\rho}(x)}{\partial x}\bigg]+o(\varepsilon^{1/2}).
\end{multline*}
Assuming $L_\varepsilon=\sqrt{(2\pi)^n/(\varepsilon |\tilde{\theta}|)}>0$, the Wick rotated Schr\"odinger type equation is,
\begin{multline*}
\Psi_s^{\tau,\rho}(x)+\varepsilon  \frac{\partial \Psi_s^{\tau,\rho}(x)}{\partial s}+o(\varepsilon)=\left\{1-\varepsilon f^\rho[s,x(\tau),u_\rho(s),u_{-\rho}^*(s)]+\mbox{$\frac{1}{2}$}\varepsilon \widehat{w}^T\tilde{\theta}^{-1}\widehat{w}\right\}\\ \times\bigg[\Psi_s^{\tau,\rho}(x)+\left[x(\tau)+\mbox{$\frac{1}{2}$} (\tilde{\theta}^{-1} \widehat{w})\right]\frac{\partial \Psi_s^{\tau,\rho}(x)}{\partial x}\bigg]+o(\varepsilon^{1/2}).
\end{multline*}
For any finite positive number $\eta$ we know $x(\tau)\leq\eta\varepsilon|\xi^T|^{-1}$, and there exists a $|\theta^{-1}w|\leq 2 \eta\varepsilon|1-\xi^T|^{-1}$ such that for $\varepsilon\ra 0$ we have $\big|x(\tau)+\mbox{$\frac{1}{2}$}(\theta^{-1} w)\big|\leq\eta\varepsilon$. Hence, 
\begin{align*}
\frac{\partial \Psi_s^{\tau,\rho}(x)}{\partial s}&=- f^\rho[s,x(\tau),u_\rho(s),u_{-\rho}^*(s)]\ \Psi_s^{\tau,\rho}(x), 
\end{align*}
with the Wheeler-Di Witt type equation is,
\begin{align}
-\frac{\partial f^\rho[s,x(\tau),u_\rho(s),u_{-\rho}^*(s)]}{\partial u_\rho}\ \Psi_s^{\tau,\rho}(x)=0,\notag
\end{align}
whose solution with respect to $u_\rho$ gives $\rho^{th}$ firm's non-cooperative feedback Nash equilibrium strategy\\ $\phi_N^{\rho*}[s,x^*(s),u_\rho^*(s),u_{-\rho}^*(s)]$.
$\square$

 \section{Discussion}
 In this paper we use a Feynman type path integral method to find optimal strategies for dynamic profit functions quadratic in time with a stochastic differential market dynamics for infinite dimensional vector spaces (i.e. Walrasian equilibrium) and finite dimensional vector spaces (i.e. Pareto optimality and Nash equilibrium). In Proposition \ref{p2} we show in the generalized non-linear case like the Merton-Garman-Hamiltonian \citep{belal2007,merton1973} Equation we are able find an optimal strategy where traditional Pontryagin's maximum principle does not work. Furthermore, in Example \ref{ex2}  where both the profit function and market dynamics are linear to strategy we are still able to find optimal strategy of a firm. Again in this case, we cannot use Pontryagin's maximum principle because after doing differentiation with respect to control, the strategy term vanishes and optimal strategy cannot be found. According to the Generalized Weierstrass Theorem we know solution exists when both the objective function and market dynamics are linear in terms of control \citep{intriligator1971}. Under Proposition \ref{pr1} we calculate a non-cooperative feedback Nash equilibrium and in the future we plan to extend this result to cooperative Nash equilibria.
 \appendix
 
 \section{Appendix section}
 
 This appendix outlines the complete assumptions required to develop the results. Throughout this paper we are considering Euclidean quantum field theory which requires further assumptions on Equation (\ref{mkt}). A quantum field is an operated valued distribution $\mathfrak{F}[s, x(s)]$ to the unbounded operators on a Hilbert space following the Garding-Wightman axioms \citep{simon1979}. Consider a measure $d\xi\equiv dx(s)$ on an Euclidean free field $\mathfrak{L}(\mathbb{R}^2)$ (The dimension is two for space-time under a Walrasian system), whose moments are the candidates of Schwinger functions. For a real valued tempered distribution $T$, let
 \begin{align}
 \left(\int\ \mathfrak{F}(y)\ f(y)\ d^2y \right)(T)&=T(f),\notag
 \end{align}
 be a random variable, where $f$ is a real valued test function of $y\in\mathbb{R}$ takes Schwinger function
 \begin{align}
 S_n(y_1,...,y_n)&=\int\ \mathfrak{F}(y_1)...\ \mathfrak{F}(y_n)\ dx\notag
 \end{align}
 which implies
 \begin{align}
 \int\ S_n(y_1,...,y_n)\ f_1(y_1)...\ f_n(y_n)\ d^{2n}y&=\int\ T(f_1)...\ T(f_n)\ dx(T).\notag
 \end{align}
 
 \begin{thm}\label{thmf}
 	[Fr\"ohlich's Reconstruction Theorem \citep{simon1979}] Let $d\xi$ be a cylindrical measure on Euclidean free field $\mathfrak{L}(\mathbb{R}^2)$ obeying the following properties:\\
 	(i) The measure $d\xi$ is invariant with respect to proper Euclidean motions of the form $T(x)\mapsto T(Ex+h)$, where $ h\in\mathbb{R}^2$ and $ E\in \boldsymbol{SO}(2)$, where $\boldsymbol{SO}(2)$ represents Lie Special orthogonal group of dimension $2$.\\
 	(ii) Osterwalder-Schrader positivity: For a given real valued test function $f$ in Garding-Wightman field $\mathfrak{L}'(\mathbb{R})$ or $f\in \mathfrak{L}'(\mathbb{R})$ with $\text{support}\ f\subset\{(s,x),\ s>0\}$, let $(\theta f)(s,x)=f(-s,x)$, where $\theta$ is a parameter. Then for real valued $f_1, f_2,...,\ f_n$ with the above support and for the set of complex numbers $z_1, z_2,..., z_n\in \mathbb{C}$ we have
 	\begin{align}
 	\sum_{j=1}^n\ \sum_{k=1}^n\ \overline{z}_kz_j\ \int\ \exp\left\{\mathfrak{i}[\mathfrak{F}(f_k)-\mathfrak{F}(\theta f_j)]\right\} d\xi\geq 0.\notag 
 	\end{align} \\
 	(iii) For any real valued test function in Euclidean free field $f\in\mathfrak{L}(\mathbb{R}^2)$, $$\int\ \exp[\mathfrak{F}(f)]\ d\xi<\infty,$$ and the action of the translations $[s,x(s)]\rightarrow[s+\varepsilon,x(s+\varepsilon)]$ is ergodic.
 \end{thm}

Assume Equation (\ref{mkt}) is in Euclidean free field and it satisfies three conditions in above Theorem \ref{thmf}. Hence, the measure $d\xi$ is cylindrical and the feasible set of Equation (\ref{mkt}) satisfies
\begin{align}\label{w3}
dx(s)\geq \mu[s,x(s),u(s)] ds+\sigma[s,x(s),u(s)] dB(s).
\end{align}
As $G[s,x(s),u(s)]=dx(s)- \mu[s,x(s),u(s)]\ ds-\sigma[s,x(s),u(s)]\ dB(s)$, Equation (\ref{w3}) implies $G[s,x(s),u(s)]\geq 0$. The
dynamic Walrasian system then satisfies
\begin{align}
&\max_{u \in U}\ \Pi(u,t)=\max_{u \in U}\ \E\ \int_{0}^{t}\ \pi[s,x(s),u(s)]\ ds,\notag
\end{align} 
with constraint $dx(s)=\mu[s,x(s),u(s)]\ ds+\sigma[s,x(s),u(s)]\ dB(s)$, and initial condition $x(0)=x_0$. Following  \cite{chow1996} at time $s'\in[0,t']$, the stochastic Lagrangian function is 
\begin{align}\label{w4}
&\int_0^{t'} \E_{s'}\biggr\{{\pi}[s',x(s'),u(s')]-\lambda \tilde{G}[s',x(s'),u(s')]\biggr\} ds',
\end{align}
where $\lambda$ is the non-negative Lagrangian multiplier, $$\tilde{G}[s',x(s'),u(s')]\ ds'={G}[s',x(s'),u(s')],$$ and $\E_{s'}$ is the conditional expectation on time $s'$, $\E_{s'}(.)=\E[.|x(s')]$. As we are interested in a forward looking process, at time $s'$ only the information up to $s'$ is available, and based on this we forecast the state of the system at time $s'+ds'$. Furthermore, in the path integral approach, Equation (\ref{w4}) corresponds to the Lagrangian function of the Feynman action functional in Minkowski space-time with imaginary time $s'$. In order to get a Euclidean path integral we need to perform the Wick rotation on Equation (\ref{w4}). Suppose, there exists dynamic non-negative measurable profit function $\hat{\pi}$ such that ${\pi}[s',x(s'),u(s')]=d^2 \hat{\pi}[s',x(s'),u(s')]/(ds')^2$. For imaginary time $s'$, the Feynman action functional becomes
\begin{align}\label{w5}
\mathcal{A}_{0,t'}^F(x)&= \int_0^{t'} \E_{s'}\biggr\{{\pi}[s',x(s'),u(s')]-\lambda \tilde{G}[s',x(s'),u(s')]\biggr\} ds'. 
\end{align}
Multiplying both sides of Equation (\ref{w5}) by $\mathfrak{i}$ and substituting $s'=-\mathfrak{i} s$ (so that $ds'=-\mathfrak{i}\ ds$) yields,
\begin{align}\label{w6}
\mathfrak{i} \mathcal{A}_{0,t'}^F(x)&=\mathfrak{i}  \int_0^{t} \E_{s}\biggr\{\left(\frac{d}{-\mathfrak{i} ds}\right)^2\hat{\pi}[s,x(s),u(s)]-\lambda \tilde{G}[s,x(s),u(s)]\biggr\} (-\mathfrak{i} ds)
\end{align}
so that
\begin{align}
\mathcal{A}_{0,t}(x)&=  \int_0^t \E_{s}\biggr\{{\pi}[s,x(s),u(s)]+\lambda \tilde{G}[s,x(s),u(s)]\biggr\} ds.\notag
\end{align}
In Equation (\ref{w6}), $\mathcal{A}_{0,t}(x)$ is defined as Euclidean action functional after the Wick rotation. Theorem \ref{thmf} and Condition (\ref{w3}) imply that if $G[s,x(s),u(s)]\geq 0$ then $\tilde{G}[s,x(s),u(s)]\geq 0$ and the parenthesis term of conditional expectation at real time $s$ is always non-negative. Now we assume some further conditions on $G[s,x(s),u(s)]$.

\begin{as}\label{as1}
	Suppose $G[s,x(s),u(s)]$ is a non-negative real valued continuous function of $(s,x,u)\in[0,t]\times X\times\mathbb{R}$ and infinitely differentiable with respect to $x$ and $u$ if $s\in[0,t]$ is fixed and $\a^{th}$ order derivatives $\partial_x^\a G[s,x,u]$ and $\partial_u^\a G[s,x,u]$ respectively are continuous functions of $(s,x,u)$ for any $\a$. Moreover, for any integer $m\geq 2$ there exist positive constants $v_{m}^1$ and $v_{m}^2$ such that, $
	\big|\partial_x^\a G[s,x,u]\big|\leq v_{ m}^1$,  $\big|\partial_u^\a G[s,x,u]\big|\leq v_{m}^2,$
	if $\a$ is an integer with $2\leq\a\leq m$ and $(s,x,u)\in[0,t]\times X\times\mathbb{R}$.
\end{as}

From Assumption \ref{as1} there exists positive constants $ v_{\mathfrak 0}^1$,  $v_{\mathfrak 1}^1$ and $v_{\mathfrak 1}^2$ such that  for all $(s,x,u)\in[0,t]\times X\times\mathbb{R}$, $\big| G(s,x,u)\big|\leq v_{\mathfrak 0}^1(1+|x|)^2$, $ \big|\partial_x G(s,x,u)\big|\leq v_{\mathfrak 1}^1(1+|x|)$ and, $ \big|\partial_u G(s,x,u)\big|\leq v_{\mathfrak 1}^2(1+|x|)$.

\begin{defn}\label{d1}
	For all $x\in[0,1]$ define a Wick rotated wave integral $I(\Psi)$ with Euclidean action function $\mathcal{A}(x)$ such that 
	\begin{align}
	I(\Psi)=\int_{\mathbb{R}}\ \exp\{-\mathcal{A}(x)\}\ \Psi(x)\ dx,\notag 
	\end{align}
	where $\Psi(x)$ is a real valued wave function of $x$. 
\end{defn}
The integration defined in Definition \ref{d1} may not converge absolutely, and we need following definition \citep{fujiwara2017}.

\begin{defn}\label{d2}
	For $\varepsilon>0$ consider a family of $C^{\infty}$, $\omega_{\varepsilon}(x)$ which follows the properties given in Definition 3.1 of \cite{fujiwara2017}. The Wick rotated wave integral is
	\begin{align}
	I(\Psi)=\lim_{\varepsilon\ra 0}\ \int_{\mathbb{R}} \omega_\varepsilon \exp\{-\mathcal{A}(x)\}\ \Psi(x) dx, \notag
	\end{align}	
	as long as\\
	(i) For any family of $\omega_{\varepsilon}(x)$ the integral $I(\omega_\varepsilon)$ converges absolutely and,\\
	(ii) The right hand side limit of Equation (\ref{w}) exists and independent of choice of $\{\omega_\varepsilon\}$. 
\end{defn} 
After using Proposition $3.1$ in \cite{fujiwara2017} and Definition \ref{d2} we conclude integral $I(\Psi)$ in Definition \ref{d1} is absolutely convergent.

\begin{as}\label{as2}
	Suppose, $x\in X$ such that;\\
	(i) The Euclidean action $\mathcal{A}(x)$ is a $C^\infty$ function. If $|\a|\geq 1$, then there exists a positive constant $\mathcal{C}_\a$ such that,
	\begin{align}
	\big|\partial_x^\a \mathcal{A}(x)\big|\leq\mathcal{C}_\a.\notag
	\end{align}
	(ii) The wave function $\Psi(x)$ which depends on $x$ is infinitely differentiable with respect to $x$. There exists a constant $\rho\geq 0$ such that for any $\a$
	\begin{align}
	\sup_{x\in X}\ (1+|x|)^{-\rho}\ \big|\partial_x^\a \Psi(x)\big|<\infty .\notag
	\end{align}	
\end{as}

\begin{lem}\label{l1}
	[Convergence of Euclidean path integral \citep{fujiwara2017}] Consider small real time interval $[s,s+\Delta s]\subset[0,t]$ such that for some positive number $\delta>0$ we have $|\Delta s|\leq\delta$ and let $\Delta:s=s_0<s_1<...<s_J<s_{J+1}=s+\Delta s$ be an arbitrary division of interval $[s,s+\Delta s]$. Suppose $\tau_j=s_j-s_{j-1}$, $|\Delta|=\max_{1\leq j\leq j+1}\ \tau_j$ and for $x\in X$ define transition function
	\begin{align}\label{w7}
	\Psi_{0,t}(x)&=\int_{A} \exp\big[-\mathcal{A}_{0,t}(x)\big] \mathfrak{D}_x,
	\end{align}
	where $A$ is the space of all paths that connect $x(0)$ to $x(t)$ and $\mathfrak{D}_x$ is a uniform measure on the space $A$. Let us define a local transition function in the interval $[s,s+\Delta]$ such that
	\begin{align}\label{w8}
	\Psi_{s,s+\Delta s}(x):=\frac{1}{L_\varepsilon} \int_{\mathbb{R}} \exp\biggr\{-\mathcal{A}_{s,s+\Delta s}(x)\biggr\} \Psi_s(x)dx
	\end{align}
	which satisfies Definitions \ref{d1} and \ref{d2} with
	\begin{align}\label{w9}
	I(\Delta,x,s,s+\Delta s):=\ \frac{1}{(L_\varepsilon)^n} \int_{\mathbb{R}^n}\ \exp\biggr\{ \sum_{j=1}^n - \mathcal{A}\big[x(s_{j-1},s_j)\big] \biggr\}\ \Psi_s(x) \prod_{j=1}^n dx(s_j),
	\end{align}
	where $\mathcal{A}_{s_{j-1},s_j}(x)$ is the Euclidean action function in $[s,s+\Delta s]$ and it is the Euclidean action function of $\tau_j$. If Equations (\ref{w7})-(\ref{w9}) satisfy Assumptions \ref{as1} and \ref{as2} then the following limit exists
	\begin{align*}
	\Psi_{0,t}(x)&=\lim_{|\Delta|\ra 0}\ I(\Delta,x,s,s+\Delta s).
	\end{align*}
\end{lem}
\bibliography{bib}
\end{document}